\documentclass[%
superscriptaddress,
 amsmath,amssymb,
 aps,
prb,
twocolumn
]{revtex4-2}

\usepackage{graphicx}
\graphicspath{{./Figs/}}

\renewcommand{\selectlanguage}[1]{}       
\DeclareUnicodeCharacter{2212}{-}         
\DeclareUnicodeCharacter{2248}{$\approx$} 
\DeclareUnicodeCharacter{FB00}{ff}        
\DeclareUnicodeCharacter{FB01}{fi}
\DeclareUnicodeCharacter{FB02}{fl}
\usepackage{dcolumn}
\usepackage{bm}
\usepackage[dvipsnames]{xcolor}

\usepackage[colorlinks=true,
            linkcolor=Blue,
            citecolor=Blue,
            urlcolor=Blue,
            pdftitle={Valence bond solids and quantum paramagnetism in the J1-J2-J3 Heisenberg model from pseudofermion functional renormalization group},
            pdfauthor={Gün Günal, Erhai Zhao, Ahmet Keleş},
            bookmarks=true,
            unicode=true
            ]{hyperref}

\usepackage{subfiles} 

\begin{document}

\preprint{APS/123-QED}

\title{Evidence of valence bond solid and quantum paramagnetism in the $J_1$--$J_2$--$J_3$ Heisenberg model from pseudofermion functional renormalization group}

\author{Gün Günal}
\email{gungunal@metu.edu.tr}
\affiliation{%
 Department of Physics, Middle East Technical University, 06800 Ankara, Turkey
}
\author{Erhai Zhao} 
\affiliation{Department of Physics and Astronomy,
  George Mason University, Fairfax, Virginia 22030,
  USA}
\author{Ahmet Keleş}%
\affiliation{%
 Department of Physics, Middle East Technical University, 06800 Ankara, Turkey
}

\date{\today}

\begin{abstract}
Motivated by recent tensor network studies reporting a quantum spin liquid (QSL) within the quantum paramagnetic (QP) regime, we investigate numerically  the zero-temperature phase diagram of the $J_1$--$J_2$--$J_3$ Heisenberg model on the square lattice using the pseudofermion functional renormalization group (pf-fRG). Thanks to improved frequency resolution and larger lattice truncations, we are able to explore a significantly larger parameter space than previous pf-fRG work to map out the phase boundary of an extended QP region. We further resolve possible bond order within the QP regime by measuring the dimer response to small deformations in $J_1$, i.e. bond anisotropy that breaks the point-group symmetry of the Hamiltonian. The functional renormalization group flow of the dimer susceptibility suggests that the QP window splits into two: a region where it grows continuously down to the lowest 
energy scale to signal valence bond solid (VBS) order, and another region where it reaches a peak at a finite scale and is subsequently suppressed to indicate a QSL. 
\end{abstract}

\maketitle

\section{\label{sec:level1}Introduction}

Understanding the interplay between frustration and quantum fluctuations in two-dimensional antiferromagnets (AFM) remains a central challenge in modern condensed matter physics. An overarching theme in the study of frustrated quantum magnets is the destabilization of long-range magnetic order, and the concurrent emergence of quantum paramagnetic (QP) states. A quantum paramagnetic regime can host various exotic phases known as quantum spin liquids (QSL) characterized by, e.g., a high degree of entanglement, fractional excitations, or even topological order~\cite{andersonResonatingValenceBonds1973,andersonResonatingValenceBond1987,balentsSpinLiquidsFrustrated2010,savaryQuantumSpinLiquids2017,broholmQuantumSpinLiquids2020}. Alternatively, this regime can harbor other non-magnetic states such as valence bond solids (VBS), where spins form singlets in repeated patterns on the lattice~\cite{readValencebondSpinPeierlsGround1989}. As these phases lack a conventional magnetic order parameter,  identifying and characterizing them using improved numerics has been an active frontier of many-body physics.

A canonical example 
is the spin-$1/2$ antiferromagnetic Heisenberg model with nearest neighbor exchange $J_1$ and next-nearest neighbor exchange interaction $J_2$ on the square lattice. While the square lattice is bipartite and geometrically unfrustrated, the inclusion of next-nearest-neighbor interactions introduces strong competition between two different magnetic orders. This model has been studied extensively and is predicted to be magnetically ordered for $J_2/J_1 \lesssim 0.4$ and $J_2/J_1 \gtrsim 0.65$, with the precise boundaries varying from one method to another. In the intermediate regime, magnetic ordering is suppressed, giving rise to a QP regime whose nature is still under debate \cite{qianAbsenceSpinLiquid2024,huangPlaquettetypeValenceBond2024,liuGaplessQuantumSpin2022,wangCriticalLevelCrossings2018,jiangSpinLiquidGround2012,gongPlaquetteOrderedPhase2014,richterSpin12J1J22010,reutherJ1J2Frustrated2010}. 

The $J_1$--$J_2$--$J_3$ Heisenberg model extends the $J_1$--$J_2$ model above by introducing a tunable third-nearest-neighbor interaction $J_3$, 
giving rise to an even richer phase diagram.
Its Hamiltonian is given by:
\begin{equation} \label{eq:hamiltonian}
    H = J_1 \sum_{\langle i,j \rangle} \mathbf{S}_i \cdot \mathbf{S}_j + J_2 \sum_{\langle\langle i,j \rangle\rangle} \mathbf{S}_i \cdot \mathbf{S}_j + J_3 \sum_{\langle\langle\langle i,j \rangle\rangle\rangle} \mathbf{S}_i \cdot \mathbf{S}_j,
\end{equation}
where $\mathbf{S}_i$ is the spin-$1/2$ operator at site $i$, and
$\langle i,j \rangle$, $\langle\langle i,j \rangle\rangle$, and $\langle\langle\langle i,j \rangle\rangle\rangle$ denote the two sites $i$ and $j$ are nearest, next-nearest, and third-nearest neighbors on the square lattice, respectively. This model has been previously studied by a range of numerical methods~\cite{moreoIncommensurateCorrelationsFrustrated1990,chubukovFirstorderTransitionFrustrated1991,ferrerSpinliquidPhaseFrustrated1993,leungNumericalEvidenceSpinPeierls1996,capriottiLowTemperatureBrokenSymmetryPhases2004,arlegoPlaquetteOrder12008,sindzingrePhaseDiagramSpin12009,murgExploringFrustratedSpin2009,ralkoGeneralizedQuantumDimer2009,wuPhaseDiagramMagnetic2022}. Recently, Liu \textit{et al.}~\cite{liuEmergenceGaplessQuantum2022,liuTensorNetworkStudy2024} employed the Projected Entangled Pair States (PEPS) ansatz to uncover the global phase diagram of the $J_1$--$J_2$--$J_3$ model. Their study identified three magnetically ordered phases, Néel, stripe, and spiral, together with a gapless QSL and two distinct VBS orders inside the QP regime. 
It has also been investigated using the pseudofermion functional renormalization group (pf-fRG) together with series expansion (SE) and coupled-cluster method (CCM) by Reuther \textit{et al.}~\cite{reutherQuantumPhasesPlanar2011}. While these authors successfully identified the magnetically ordered phases and established the existence of a QP regime, their complementary analysis using SE and the CCM did not yield a definitive conclusion regarding the nature of the quantum paramagnetic state.
This discrepancy between PEPS and pf-fRG raises the natural question: can modern pf-fRG resolve the QP phase boundary, differentiate VBS from QSL, and serve as an independent crosscheck of the proposed PEPS phase diagram of the  $J_1$--$J_2$--$J_3$ model?

To answer this question, in this work, we revisit the phase diagram of the $J_1$--$J_2$--$J_3$ model using pf-fRG. We go beyond the previous pf-fRG study~\cite{reutherQuantumPhasesPlanar2011} by significantly broadening the parameter space to facilitate a direct comparison with recent PEPS results. Utilizing the \texttt{PFFRGSolver.jl} package~\cite{kieseMultiloopFunctionalRenormalization2022}, we employ larger lattice truncations and an improved frequency parametrization scheme to increase numerical accuracy and resolution. Furthermore, we probe incipient VBS orders such as the formation of dimer or plaquette valence bonds
with enlarged unit-cells within the QP regime to distinguish them from candidate QSL states.
Our results reveal an extended QP regime that hosts intriguing competition and interplay between candidate QSL and VBS order that had not been observed in earlier pf-fRG studies for the $J_1$--$J_2$ model~\cite{reutherJ1J2Frustrated2010}.

The remainder of this paper is organized as follows. Section~\ref{sec:level2} introduces the pf-fRG method, the observables we compute, and the bond-anisotropy probe used to detect VBS order. Sec.~\ref{sec:level3} presents our results: Sec.~\ref{sec:j1j2} benchmarks the method on the $J_1$--$J_2$ limit, Sec.~\ref{sec:global} maps the global phase diagram of magnetic instabilities, and Sec.~\ref{sec:vbs} probes the quantum paramagnetic regime using enlarged unit cells. Section~\ref{sec:level6} summarizes our conclusions. Technical details about lattice symmetries for dimer and plaquette unit cells, and the dependence of  the dimer susceptibility on $L$,
the truncation range are given in the supplementary materials \cite{supplementary_materials}.

\section{\label{sec:level2}Pseudofermion Functional Renormalization Group}

Pseudofermion functional Renormalization Group (pf-fRG) is a versatile and powerful numerical technique to 
investigate the competing orders in quantum magnets in two- and three-dimensional lattices~\cite{reutherJ1J2Frustrated2010, mullerPseudofermionFunctionalRenormalization2024,kelesAbsenceLongrangeOrder2018,kelesRenormalizationGroupAnalysis2018,kelesRiseFallPlaquette2022,iqbalIntertwinedNematicOrders2016,reutherQuantumPhasesPlanar2011,reutherFunctionalRenormalizationGroup2011,gresistaUnconventionalOrdersMapleleaf2025,gottelCriticalScalesAnisotropic2012,iqbalSpinLiquidNature2016,iqbalFunctionalRenormalizationGroup2016}.
We follow the approach pioneered in \cite{reutherJ1J2Frustrated2010} and focus on zero temperature.
It begins by representing spin operators as pseudofermions which allow the use of field-theoretical techniques including functional Renormalization Group for interacting fermions. The Renormalization Group (RG) flow is generated by introducing a sliding frequency scale $\Lambda$, e.g. by using a function $\Theta(\Lambda)$ as a cutoff to select higher frequency modes with $\Omega>\Lambda$ in the Green's function.
The resulting infinite hierarchy of differential equations, relating the $n$-particle to the $(n+1)$-particle correlators, are made tractable by truncating the hierarchy at the six-point level using the Katanin truncation scheme~\cite{kataninFulfillmentWardIdentities2004}.

The flow is initiated at a large ultraviolet scale $\Lambda\to\infty$, where the interaction vertices are determined by the bare Hamiltonian, and proceeds towards the infrared $\Lambda\to0$. Quantum fluctuations at scale $\Omega>\Lambda$ are successively integrated out during the flow, so the full Green's function and interaction vertices may be analyzed to detect the emergence of incipient order (or the lack thereof) at some low energy scale.
More details can be found in Refs. \cite{kelesAbsenceLongrangeOrder2018,kelesRenormalizationGroupAnalysis2018,kelesRiseFallPlaquette2022}. 
The main advantage of pf-fRG is its capability of treating all instabilities, or orders, on equal footing without a prior bias. Another advantage is that since it is formulated in real space, given sufficient computing resources, one can access large clusters to accurately describe spin systems with longer range interactions, e.g. decaying according to a power law~\cite{kelesAbsenceLongrangeOrder2018,kelesRenormalizationGroupAnalysis2018}, or with large unit cells or complex spatial profiles in the broken symmetry phase.

\begin{figure}[t]
    \centering
    \includegraphics{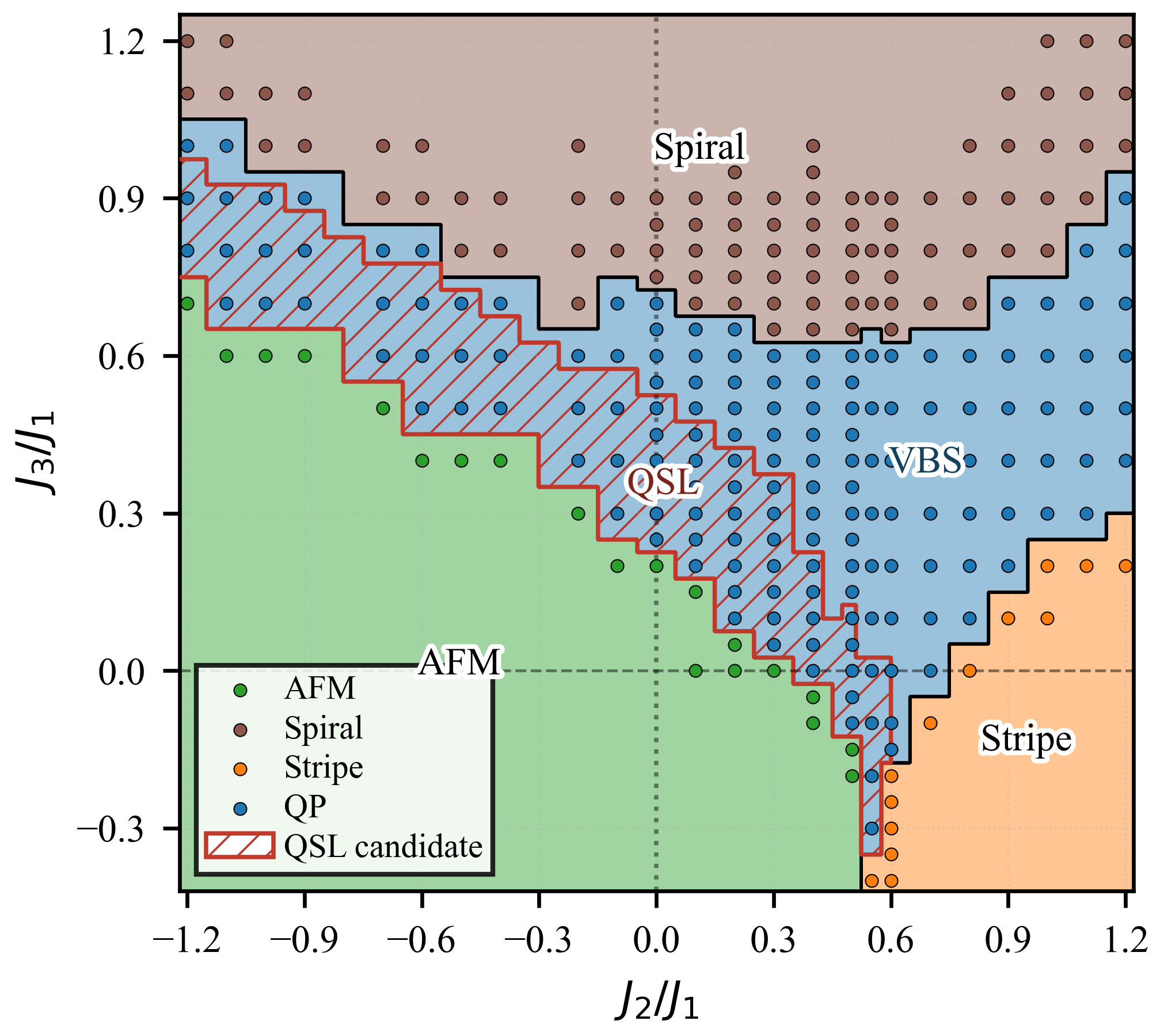}
    \caption{Phase diagram of the $J_1$--$J_2$--$J_3$ model on square lattice obtained with pf-fRG
    where each dot represents an independent calculation. The colors correspond to the identified phases: antiferromagnetic (AFM) order in green, stripe order in orange, spiral order in brown, and the quantum paramagnetic (QP) regime in blue. We distinguish two different QP phases: candiate qauntum spin liquid (QSL) and valence bond solid (VBS), separated with red dashed area, by monitoring the  
    dimer susceptibility flows 
    (see main text). }
    \label{fig:phase_diagram}
\end{figure}

We solve these flow equations using the open-source \texttt{PFFRGSolver.jl} Julia package~\cite{kieseMultiloopFunctionalRenormalization2022}. This makes it much easier to reproduce the calculations reported here. For spin models with further neighbor interactions and complex symmetry breaking patterns, it is crucial to have sufficient  frequency resolution.
The Julia package employs adaptive frequency grids combined with asymptotic classes, which utilize the high-frequency decaying parts of the Green's functions. These features turn out to be crucial if one aims to going beyond previous results to obtain a more complete and more refined phase diagram such as Fig.~\ref{fig:phase_diagram}. Details regarding the asymptotic classes~\cite{wentzellHighfrequencyAsymptoticsVertex2020} and their application to pf-fRG can be found in Refs.~\cite{kieseMultiloopFunctionalRenormalization2022,thoennissMultiloopPseudofermionFunctional2020}.

While pf-fRG can be formulated in the real space for 
an infinite lattice, any practical numerical implementation requires truncation, i.e., only keeping the spin-spin correlation functions, e.g. between site $i$ and site $j$,  up to some distance cutoff. In our calculations, we adopt a spatial truncation range of $L=11$ lattice spacings, discarding any correlations beyond this distance, i.e., $|j-i|>L$. For selected points in the parameter space, we increase this cutoff up to $L=25$. Furthermore, we exploit the lattice symmetries of the Hamiltonian to reduce the number of vertices. By default, our calculations exploit the full point-group symmetry, $C_{4}$, of the unperturbed square lattice $J_1$--$J_2$--$J_3$ Hamiltonian; see the Supplemental Material~\cite{supplementary_materials} for implementation details.

To detect the emergence of long-range magnetic order in pf-fRG, we employ the scale-dependent static spin susceptibility $\chi_{ij}^\Lambda$:
\begin{equation}
    \chi_{ij}^\Lambda (\Omega=0) = \int_{0}^{\infty} d\tau \, \langle T_\tau S_i^z(\tau) S_j^z(0) \rangle .
\end{equation}
During the renormalization flow, pf-fRG keeps track of the static susceptibility. An instability or a divergence of $\chi^\Lambda$ signals the onset of magnetic long-range order. To characterize this ordered state, the static susceptibility is Fourier transformed into momentum space, $\chi^\Lambda(\mathbf{q})$. The nature of the long-range order is then identified by looking at the locations of the peaks $\chi^\Lambda(\mathbf{q})$ in the Brillouin zone. 
The absence of any divergence, i.e., a smooth flow down to the lowest accessible infrared scale, indicates a paramagnetic state. 

While pf-fRG excels at identifying various magnetic instabilities, distinguishing a VBS phase (e.g. from a QSL) remains a long-standing challenge. Because the standard flow equations are truncated at the three-particle level, the higher-order correlation functions required to directly 
detect a VBS order parameter cannot be accessed within the current implementation of pf-fRG. 
Instead, we probe the system's susceptibility to VBS formation by adding a ``source field" to the Hamiltonian $H$ to induce the desired VBS order that explicitly breaks the original symmetries of $H$~\cite{reutherJ1J2Frustrated2010,iqbalIntertwinedNematicOrders2016,kiesePinchpointsHalfmoonsStars2023,kelesRiseFallPlaquette2022,iqbalSpinLiquidNature2016}.

More explicitly, we replace the uniform nearest-neighbor interaction $J_1$ in Eq. \eqref{eq:hamiltonian} with
$J'_1 = J_1 \pm \delta$, where the bond anisotropy $\delta = 0.01 J_1$ is a small perturbation. In other words, the bond 
(the strength of exchange interaction between nearest neighbors) now alternates slightly in some spatial pattern.  
The resultant perturbed Hamiltonian 
$H'$ then breaks the original $C_4$ symmetry to have a larger unit cell. We consider two kinds of bond anisotropies that correspond to the
columnar dimer and plaquette VBS. The unit cell sizes are $2\times1$ and $2\times2$, respectively.
The response to the perturbation is quantified by the dimer susceptibility~\cite{reutherJ1J2Frustrated2010}:
\begin{equation}\label{eq:order}
    \chi^\Lambda_{p,d} = \frac{J_1}{\delta} \frac{\chi_s^\Lambda - \chi_w^\Lambda}{\chi_s^\Lambda + \chi_w^\Lambda} \ ,
\end{equation}
where $\chi_s^\Lambda$ and $\chi_w^\Lambda$ denote the nearest-neighbor static spin susceptibilities with the strengthened and weakened bonds, respectively. The subscripts $p$ and $d$ distinguish the response calculated using the $2\times2$ plaquette and $2\times1$ dimer unit cells. By construction, the flow starts at $\chi^{\Lambda \to \infty}_{p,d}=1$. We take a significant enhancement above this initial value as evidence for a VBS ground state~\cite{mullerPseudofermionFunctionalRenormalization2024,reutherJ1J2Frustrated2010}. Details about enlarged unit cells and the corresponding symmetries can be found in 
\cite{supplementary_materials}.
This approach has been applied successfully to detect the plaquette order in the Shastry-Sutherland model \cite{kelesRiseFallPlaquette2022}.

\section{\label{sec:level3}The fRG phase diagram}
\subsection{\label{sec:j1j2}The Limit of $J_1$--$J_2$ Model}

\begin{figure}[t]
    \centering
    \includegraphics{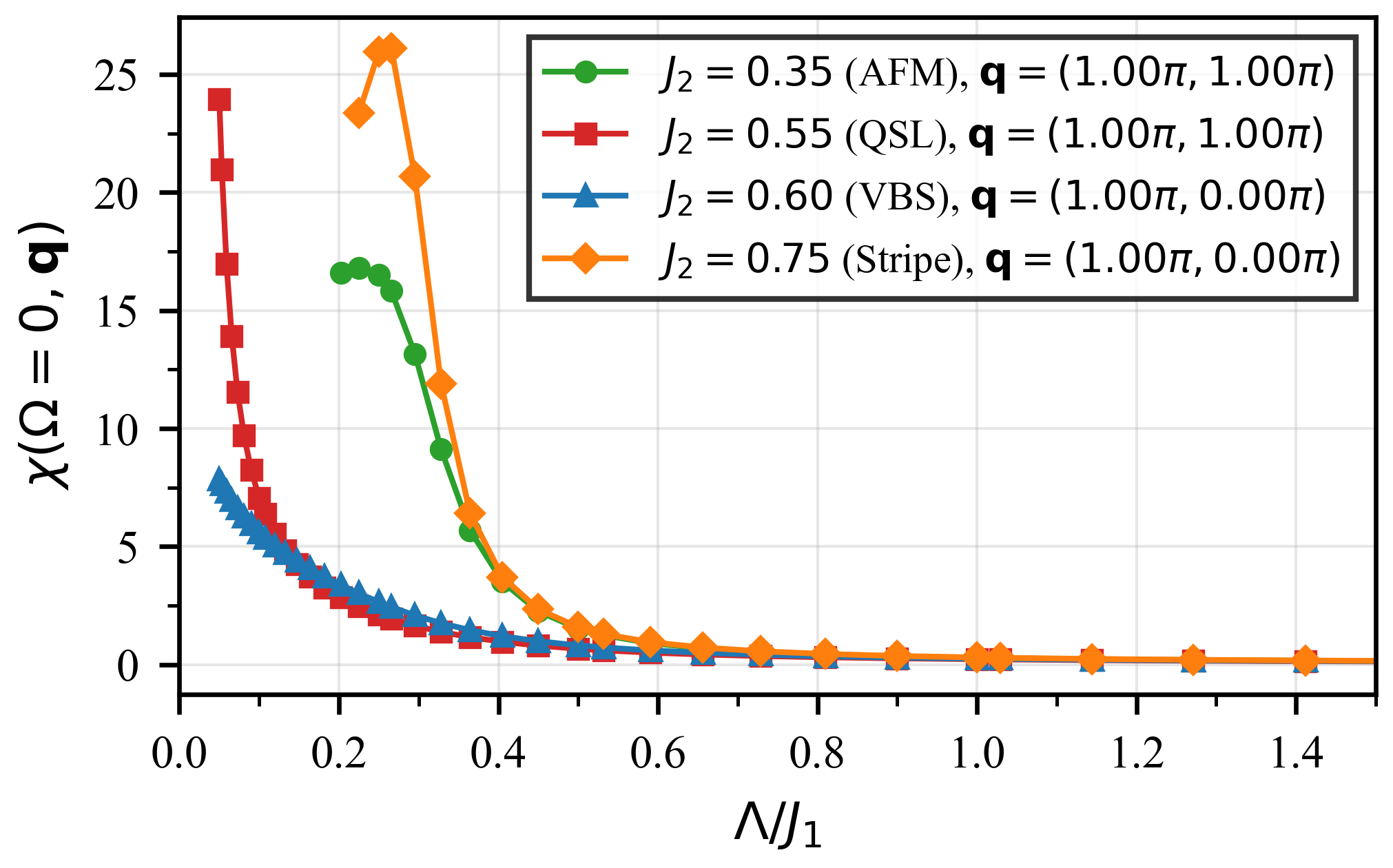}
    \includegraphics{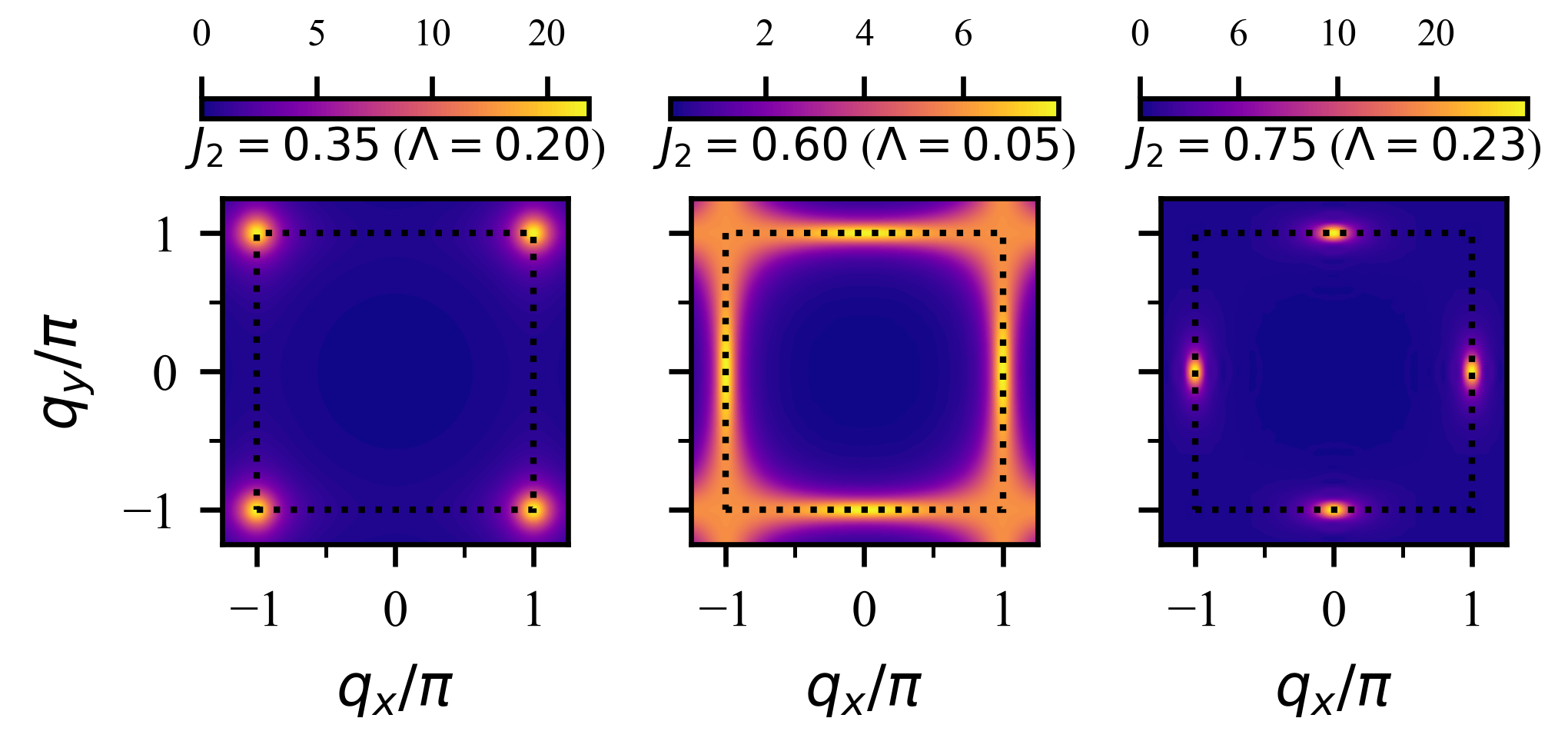}
    \caption{Static spin susceptibility in the $J_1$--$J_2$ model. Upper panel: flow of $\chi^\Lambda(\mathbf{q})$ in the leading momentum channel for representative points in the AFM phase ($J_2/J_1=0.35$, green), the candidate QSL regime ($J_2/J_1=0.55$, red), the VBS regime ($J_2/J_1=0.60$, blue), and the stripe phase ($J_2/J_1=0.75$, orange). The magnetically ordered phases exhibit an instability that terminates the flow at a critical scale $\Lambda_c$, whereas within the QP window the susceptibility flows smoothly down to the lowest scale $\Lambda=0.05 J_1$. Lower panel: the corresponding momentum-resolved susceptibility $\chi^\Lambda(\mathbf{q})$ evaluated at the end of the flow.}
    \label{fig:J300_chiflow}
\end{figure}
\begin{figure}[t]
    \centering
    \includegraphics{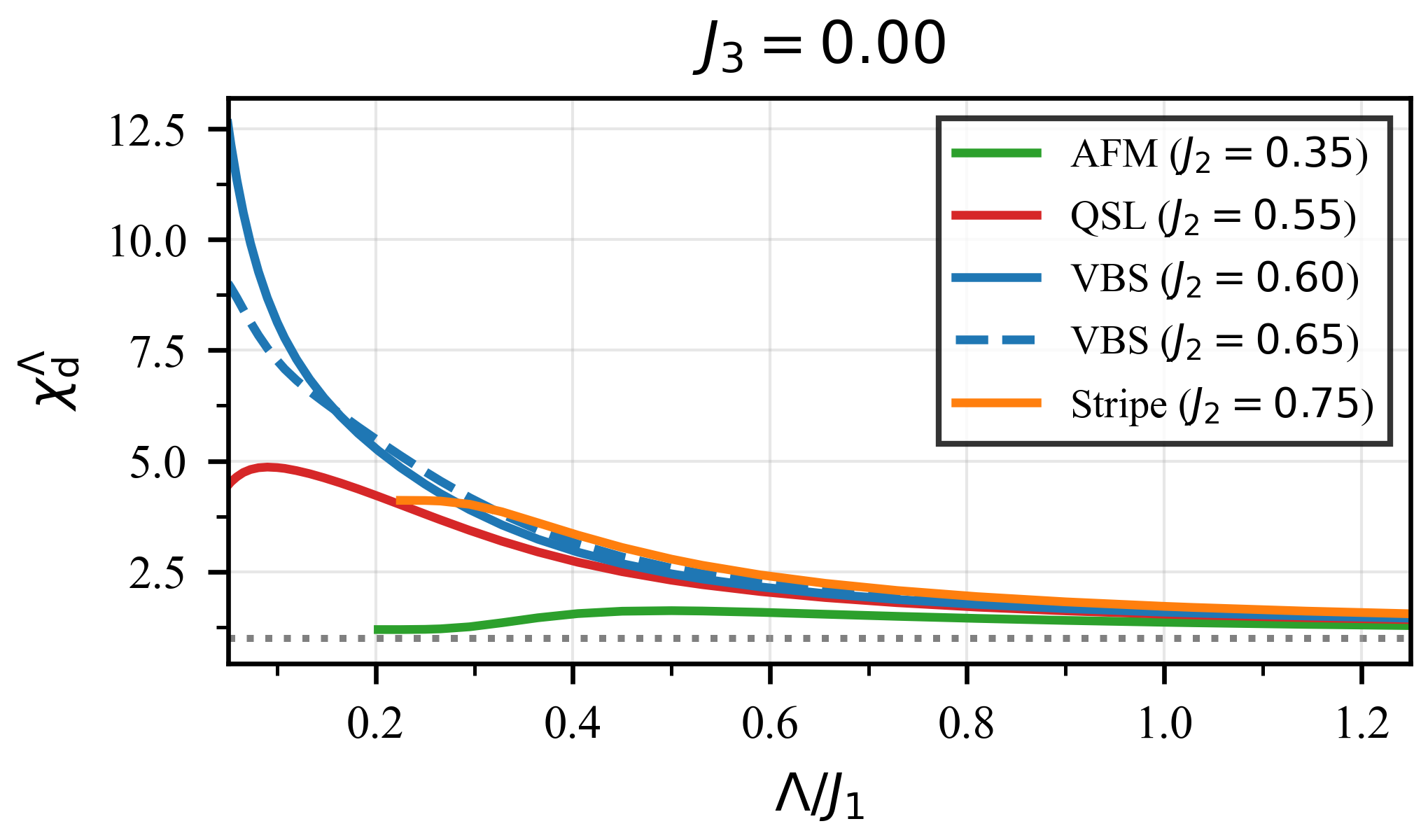}
    \caption{Renormalization flow of the dimer susceptibility $\chi_d^\Lambda$ in the $J_1$--$J_2$ model for selected values of $J_2$. The horizontal gray dotted line marks the unperturbed initial value $\chi_d=1$. In the AFM phase ($J_2/J_1=0.35$, green) the perturbation is suppressed before the flow breaks down. In the frustrated regime ($J_2/J_1=0.55$, red), $\chi_d^\Lambda$ grows initially but is ultimately suppressed at low energy scales. In the VBS regime, it instead displays a significant enhancement down to the lowest scales smoothly, ($J_2/J_1=0.60$ and $0.65$, solid and dashed blue, respectively). 
    In the stripe phase ($J_2/J_1=0.75$, orange), the dimer susceptibility plateaus and the magnetic instability terminates the flow.}
    \label{fig:J3_000_VBS_flow}
\end{figure}

We begin with the special case of $J_3=0$ where $H$ reduces to the well-known antiferromagnetic $J_1$--$J_2$ Heisenberg model on the square lattice. This limit corresponds to a horizontal cut (dashed line) with $J_3=0$ in Fig.\ref{fig:phase_diagram}.
Despite several decades of intense study, the nature of the frustrated regime of the $J_1$--$J_2$ model remains controversial.  Consequently, it has served as a central testing ground for competing numerical schemes to claim the state of the art~\cite{qianAbsenceSpinLiquid2024,huangPlaquettetypeValenceBond2024,liuGaplessQuantumSpin2022,wangCriticalLevelCrossings2018,jiangSpinLiquidGround2012,gongPlaquetteOrderedPhase2014,richterSpin12J1J22010,reutherJ1J2Frustrated2010}. Here we use this limit to benchmark our pf-fRG implementation and to establish the flow signatures on which the analysis of the full parameter space in Secs.~\ref{sec:global} and~\ref{sec:vbs} relies. We shall see that distinctive flow behaviors emerge at different values of $J_2/J_1$.

The upper panel of Fig.~\ref{fig:J300_chiflow} shows the flow of the static susceptibility in the leading momentum channel (i.e.
the maximum of $\chi_d^\Lambda(\mathbf{q})$ in the momentum space) as a function of $\Lambda$, the sliding RG scale. The onset of magnetic long-range order is signaled by a breakdown of this flow at a critical scale $\Lambda_c$. The ordering pattern is then identified by resolving the peak locations of $\chi^\Lambda(\mathbf{q})$ in the momentum space, e.g. $(\pi,\pi)$ versus $(\pi,0)$, near the end of the flow, as shown in the lower panel of Fig.~\ref{fig:J300_chiflow}.

For $J_2/J_1 \lesssim 0.35$ the flow becomes unstable (green curve) and $\chi^\Lambda(\mathbf{q})$ develops sharply localized peaks at the corners of the Brillouin zone, $\mathbf{q}=(\pi,\pi)$, identifying the Néel state, which we refer to throughout as the AFM phase. Above this value, e.g., when $J_2/J_1=0.55$, the instability disappears and the susceptibility flows smoothly down to the lowest scale $\Lambda=0.05 J_1$, while the peaks in $\chi^\Lambda(\mathbf{q})$ smear out, signaling the lack of long-range magnetic order to indicate a quantum paramagnetic phase. Within this QP window the peaks move continuously from $(\pi,\pi)$ towards $(\pi,0)$, after which they sharpen again upon approaching the opposite boundary, in agreement with the earlier observation of Reuther and Wölfle~\cite{reutherJ1J2Frustrated2010}. The flow becomes unstable once more at $J_2/J_1 \simeq 0.75$, where the peak has localized at $\mathbf{q}=(\pi,0)$, marking the onset of stripe order. The QP regime obtained in this way therefore extends over approximately $0.36 \lesssim J_2/J_1 \lesssim 0.74$. Note that, in the case of improved frequency resolution employed here, by ``divergence'' of the flow we refer to finite-size  behavior, not a numerical divergence. In the long-range order phases, the susceptibility peak grows systematically with system size as shown in \cite{supplementary_materials}, which is consistent with a divergence in the thermodynamic limit. 

A closer look at the intermediate region via the dimer susceptibility $\chi_d^\Lambda$ reveals that the 
QP window consists of two distinct regimes, and their differences are highlighted in Fig.~\ref{fig:J3_000_VBS_flow}. For $0.36 \lesssim J_2/J_1 \lesssim 0.57$, during the flow $\chi_d^\Lambda$ slowly rises to a broad local maximum at a certain scale $\Lambda^*$, but subsequently it is suppressed as the flow proceeds (red curve, $J_2/J_1=0.55$). We interpret this downward turn at small $\Lambda$ as an indication of the system rejecting the imposed bond perturbation to disfavor a VBS ground state~\cite{reutherJ1J2Frustrated2010,mullerPseudofermionFunctionalRenormalization2024}, leaving open the possibility of a QSL region. A similar, but milder, flow pattern is observed in the neighboring AFM phase (green curve, $J_2/J_1=0.35$), where $\chi_d^\Lambda$ eventually decreases before the magnetic instability terminates the flow. For $J_2/J_1 \gtrsim 0.57$ no such maximum develops. Instead, $\chi_d^\Lambda$ grows monotonically down to the lowest RG scale (solid blue curve, $J_2/J_1=0.60$), which we interpret as the onset of VBS order. Across this region the susceptibility reaches final values that are $5$ to $15$ times the unperturbed one. The response is strongest near $J_2/J_1 \simeq 0.60$ and then weakens (dashed blue curve, $J_2/J_1=0.65$) gradually as the stripe phase is approached. Once inside the stripe phase, $\chi_d^\Lambda$ saturates at values much smaller than those reached in the VBS regime, and the flow is terminated by the magnetic instability (orange curve, $J_2/J_1=0.75$). We have verified that the phase boundaries are unchanged upon further increase in $L$, the truncation range.

To summarize, along the cut $J_3=0$, pf-fRG yields the phase sequence AFM, (candidate) QSL, VBS, and finally stripe order as $J_2/J_1$ is increased. Our results here thus reproduces the sequence proposed by several previous studies of the model~\cite{liuGaplessQuantumSpin2022,gongPlaquetteOrderedPhase2014,huangPlaquettetypeValenceBond2024,wangCriticalLevelCrossings2018}, whose precise phase diagram nevertheless remains a matter of debate. The study of the cut $J_3=0$ gives us the confidence about our pf-fRG implementation, including its pros and cons. It is also illuminating to compare our results to those obtained by Reuther and Wölfle in their original pf-fRG study of the same model~\cite{reutherJ1J2Frustrated2010}. These authors introduced the probe based on dimer susceptibility described above, and they found $\chi_d^\Lambda$ and $\chi_p^\Lambda$ are enhanced at the end of the flow but without developing a divergence, leaving open whether the paramagnet is a spin liquid with strong dimer fluctuations or a genuine VBS. Our conclusion is more refined: we are able to resolve two distinct flow patterns within the QP window. We attribute the new capability to the improved frequency parametrization and the larger spatial truncation ranges employed in our numerical implementation, both of which were identified in Ref.~\cite{reutherJ1J2Frustrated2010} as the factors limiting the resolution at low $\Lambda$. On the other hand, we do confirm their observation that the $2\times1$ and $2\times2$ unit cells yield identical responses. Resolving the underlying VBS order thus remains an open problem for pf-fRG.

The QP window obtained here is wider than the range $0.4 \lesssim J_2/J_1 \lesssim 0.65$ reported in the literature on the basis of a variety of methods, including DMRG, tensor-network, Monte Carlo, and neural-network approaches~\cite{qianAbsenceSpinLiquid2024,huangPlaquettetypeValenceBond2024,liuGaplessQuantumSpin2022,wangCriticalLevelCrossings2018,jiangSpinLiquidGround2012,gongPlaquetteOrderedPhase2014,richterSpin12J1J22010}. The overestimation of the QP regime is a known feature of pf-fRG, which  does not account for all quantum fluctuations due to truncations~\cite{mullerPseudofermionFunctionalRenormalization2024}. Compared to these other methods, the main advantages of pf-fRG are its relatively lower computational cost and its capability of reaching large system sizes ($L$) and higher energy and momentum resolutions without any prior bias. They make it feasible to carry out dense parameter scans to be presented in the following sections.

\subsection{\label{sec:global}Identifying the Magnetic Instabilities and Quantum Paramagnetism}

We construct the global phase diagram of the $J_1$--$J_2$--$J_3$ model by 
a systematic scan of the parameter space, taking multiple cuts along the $J_2$ and $J_3$ axes with an increment of $0.1 J_1$. For a few selected cuts, a smaller increment of $0.05 J_1$ is used, so that the phase boundaries reported here are within the precision of  from $0.05$ to $0.1 J_1$. Once deep inside the phases with long-range order, which is indicated by the magnetic instability setting in at significantly higher $\Lambda_c$, we cease further stepping into these ordered phases to focus primarily on the phase boundaries and quantum paramagnetic regions.

The results are summarized in Fig.~\ref{fig:phase_diagram}, where each dot represents an independent pf-fRG calculation. (In this subsection, we focus on calculations with $C_4$ symmetry; diagnosis of possible bond border will be discussed in the next subsection). 
We observe three expected magnetic orders: AFM, stripe, and spiral, surrounding a central, extended QP regime. 
This is consistent with previous studies~\cite{capriottiLowTemperatureBrokenSymmetryPhases2004,liuEmergenceGaplessQuantum2022,liuTensorNetworkStudy2024,reutherQuantumPhasesPlanar2011}. We emphasize that compared to the earlier pf-fRG study of this model~\cite{reutherQuantumPhasesPlanar2011}, the calculations presented here cover extended parameter regimes that include negative $J_2$ and $J_3$, they are also more accurate thanks to the fact that we employ adaptive frequency grids with asymptotic classes and a larger spatial truncation range.

\begin{figure}[t]
    \centering
    \includegraphics{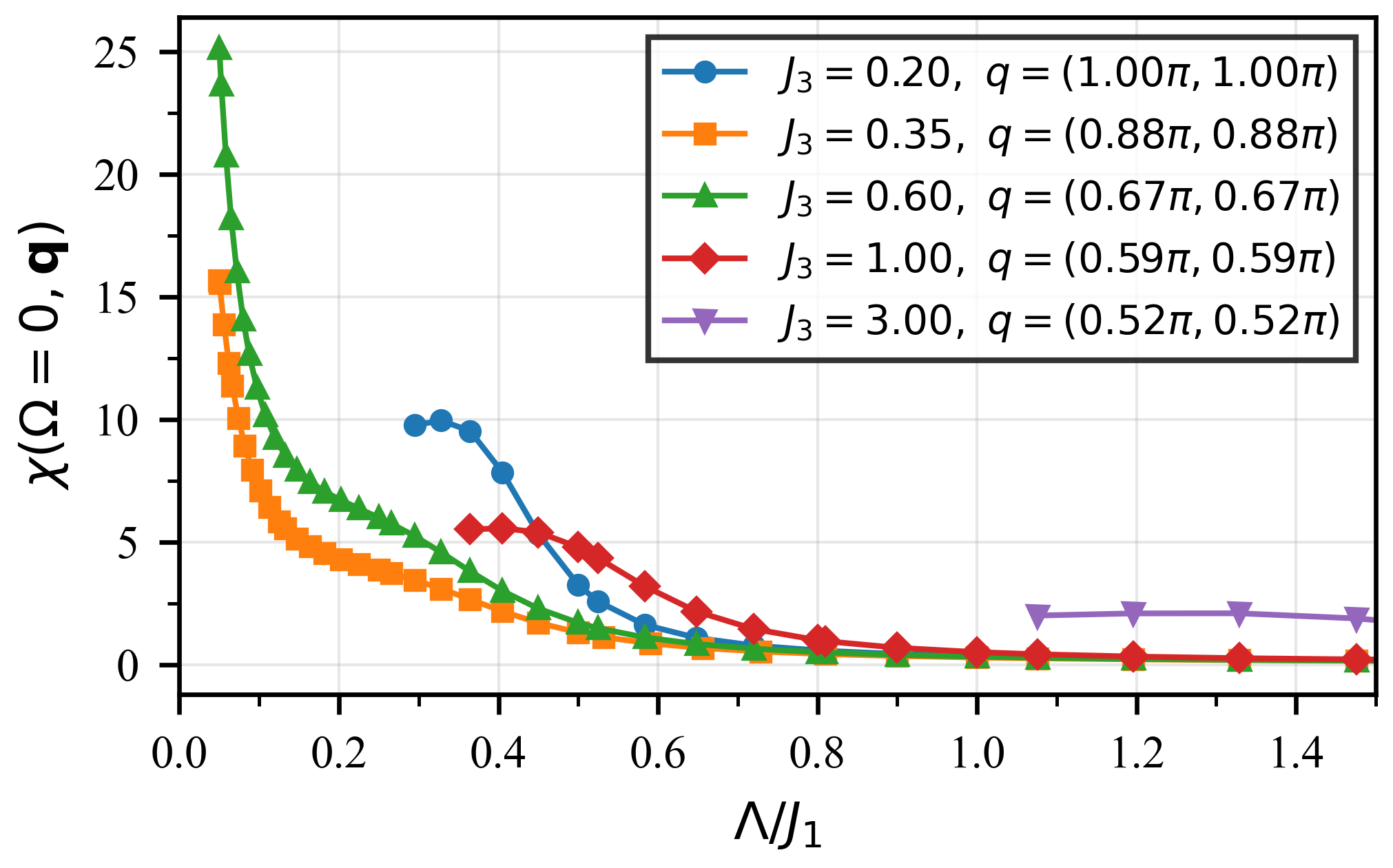}
    \includegraphics{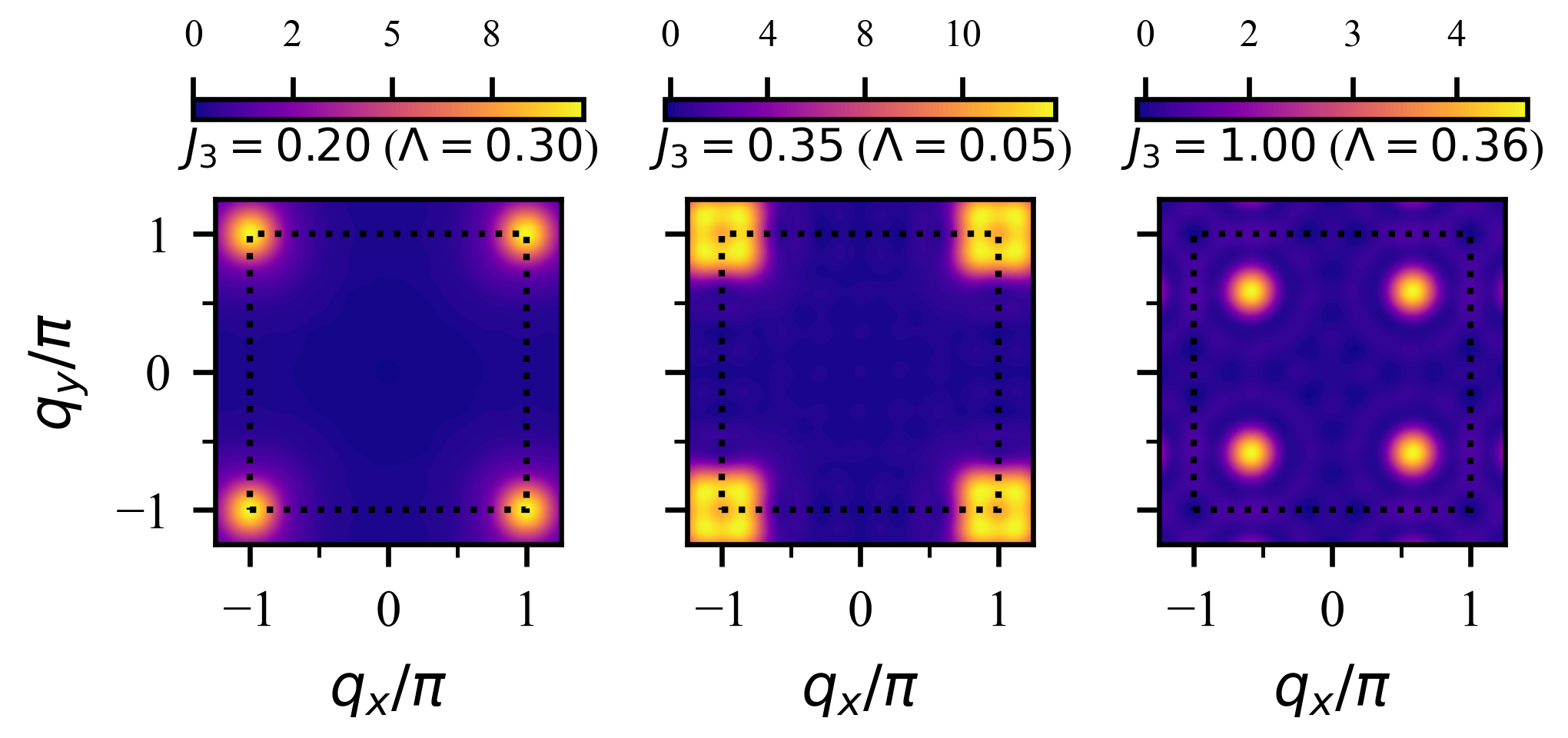}
    
    \caption{Static spin susceptibility along the $J_2=0$ cut. Upper panel: flow of $\chi^\Lambda(\mathbf{q})$ in the leading momentum channel for representative points in the AFM phase ($J_3/J_1=0.2$), the QP regime ($J_3/J_1=0.35$ and $0.6$), and the spiral phase ($J_3/J_1=1.0$ and $3.0$). The magnetically ordered phases exhibit an instability that terminates the flow at a critical scale $\Lambda_c$, whereas within the QP window the susceptibility flows smoothly down to the lowest scale $\Lambda=0.05 J_1$. Lower panel: the corresponding momentum-resolved susceptibility $\chi^\Lambda(\mathbf{q})$ for the AFM ($J_3/J_1=0.2$), the QP ($J_3/J_1=0.35$), and spiral ($J_3/J_1=1.0$) states, with the susceptibility profiles evaluated at the end of the flow.}

    \label{fig:J2_0_merged}
\end{figure}

The $J_3=0$ cut is already discussed in the previous subsection. To further illustrate how each phase is identified and the phase boundaries are drawn, we consider another cut through the phase diagram: the vertical $J_2=0$ cut, along which the Hamiltonian reduces to the $J_1$--$J_3$ model which itself is the subject of dedicated studies~\cite{leungNumericalEvidenceSpinPeierls1996,wuPhaseDiagramMagnetic2022}. The upper panel of Figure~\ref{fig:J2_0_merged} shows the flow of the static susceptibility along this cut, while the lower panel illustrates the momentum-resolved susceptibility $\chi^\Lambda(\mathbf{q})$ at three representative points. For $J_3 = 0.2 J_1$, the system resides within the AFM phase: the flow exhibits an onset of instability around $\Lambda_c \approx 0.3 J_1$ (upper panel), and the corresponding susceptibility features localized peaks at the corners of the Brillouin zone, $\mathbf{q}=(\pi, \pi)$. As $J_3$ increases, 
frustration (i.e., the competition between $J_3$ and $J_1$) suppresses the AFM order. The QP regime, e.g., at $J_3 = 0.35 J_1$, is characterized by flows that remain smooth, finite, and stable all the way down to the lowest infrared scale $\Lambda = 0.05 J_1$. Correspondingly, the 
peaks of $\chi^\Lambda(\mathbf{q})$ smear out and stray away from $(\pi, \pi)$ (see the lower panel), indicating the absence of long-range magnetic order. As $J_3$ is increased further, the peaks in the Brillouin zone start to move along the diagonal towards the center and become sharp again. At $J_3=0.84 J_1$, the flow exhibits an instability and long-range magnetic order re-emerges. The corresponding $\chi^\Lambda(\mathbf{q})$ peaks at $\mathbf{q}=(\pm q,\pm q)$, signaling a spiral order. These peaks continue to move along the  Brillouin zone diagonal towards the $(\pm \pi/2,\pm \pi/2)$ with increasing $J_3$. This progression from the AFM phase, through a disordered QP window, to spiral order 
is fully consistent with
the phase sequence obtained in recent PEPS calculations~\cite{liuEmergenceGaplessQuantum2022,liuTensorNetworkStudy2024}, with the spiral order in our case setting in at a  larger value of $J_3$. 

Our calculation also confirms the suppression of certain classical magnetic orders across the broader phase diagram. Classically, the $J_1$--$J_2$--$J_3$ model allows for two distinct spiral modes with leading momentum vectors located at $\mathbf{q}=(q,q)$ and $\mathbf{q}=(\pi,q)$~\cite{moreoIncommensurateCorrelationsFrustrated1990,chubukovFirstorderTransitionFrustrated1991,ferrerSpinliquidPhaseFrustrated1993}. However, with the inclusion of quantum fluctuations, the $(\pi,q)$ spiral order completely melts. In our pf-fRG framework, this melting is explicitly evidenced by the continuous flow of the static susceptibility at the $(\pi,q)$ wave vector down to the smallest accessible infrared scale, showing no signs of divergence. Consequently, the only stabilized spiral phase we observe is the $\mathbf{q}=(q,q)$ mode, a result that agrees with the previous pf-fRG study of Reuther \textit{et al.}~\cite{reutherQuantumPhasesPlanar2011} and recent PEPS findings~\cite{liuEmergenceGaplessQuantum2022,liuTensorNetworkStudy2024}.

While we significantly extend the parameter space explored by earlier pf-fRG studies~\cite{reutherQuantumPhasesPlanar2011} into the ferromagnetic regimes of $J_2<0$ and $J_3<0$, our phase boundaries are in excellent agreement with these earlier findings within the overlapping domains. The expanded map also allows a comparison with recent PEPS studies. 
A preliminary finite correlation radius extrapolation using $L=7, 11$ and $19$ along the cut $J_2=0.55J_1$ suggests a tricritical point in the range $J_3=-0.19J_1$ to $J_3=-0.16 J_1$.
In PEPS the non-magnetic phases of the $J_1$--$J_2$ model are far more fragile against a ferromagnetic $J_3$, disappearing already for $J_3 \lesssim -0.05 J_1$~\cite{liuTensorNetworkStudy2024}, whereas in our calculation the QP region persists down to $J_3 \approx -0.20 J_1$. On the opposite side of the phase diagram, we observe that the AFM and spiral regions remain separated by the QP regime throughout the range we scanned, whereas PEPS reports 
a direct AFM--spiral transition near $J_2 = -1.2 J_1$ where 
the paramagnetic region closes~\cite{liuTensorNetworkStudy2024}. Both differences are consistent with the tendency of pf-fRG to overestimate the paramagnetic window~\cite{mullerPseudofermionFunctionalRenormalization2024}.

To summarize, our pf-fRG scans using $H$ with $C_4$-symmetry successfully identifies the boundaries of the magnetically ordered phases, it characterizes the entire intermediate region simply as a featureless QP state. Tensor network calculations resolve this region into a gapless QSL and two distinct valence bond solids, a plaquette VBS and a mixed columnar-plaquette VBS~\cite{liuTensorNetworkStudy2024}. To detect these possible VBS orders within pf-fRG, in the next subsection, we 
consider $H'$ with weak bond anisotropies that break the symmetries of the original $J_1$--$J_2$--$J_3$ Hamiltonian.

\subsection{\label{sec:vbs}Probing Possible VBS Orders within the Paramagnetic Region}

\begin{figure}[t]
    \centering
    \includegraphics[width=\columnwidth]{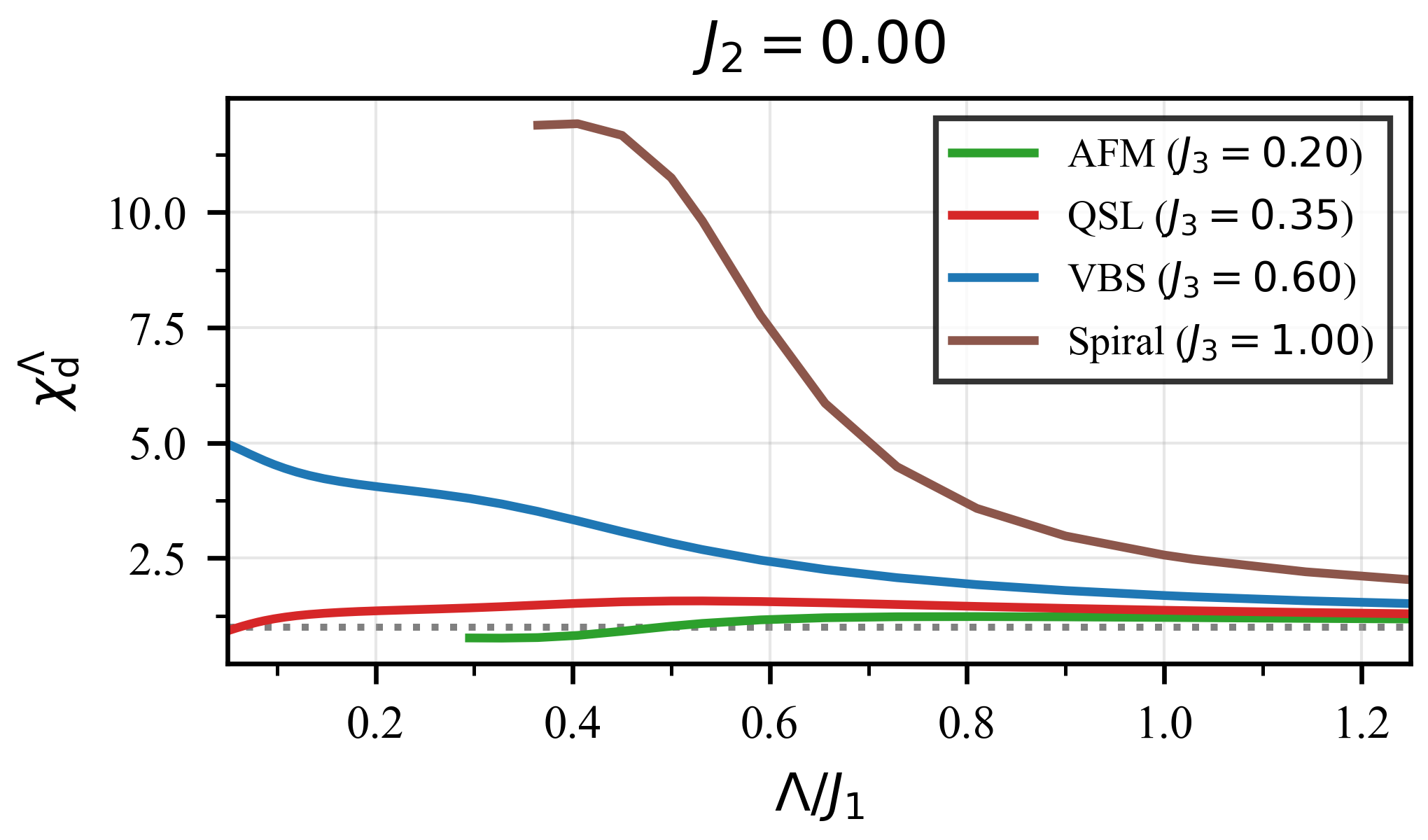}
    \vspace{-0.2cm} 
    \includegraphics[width=\columnwidth]{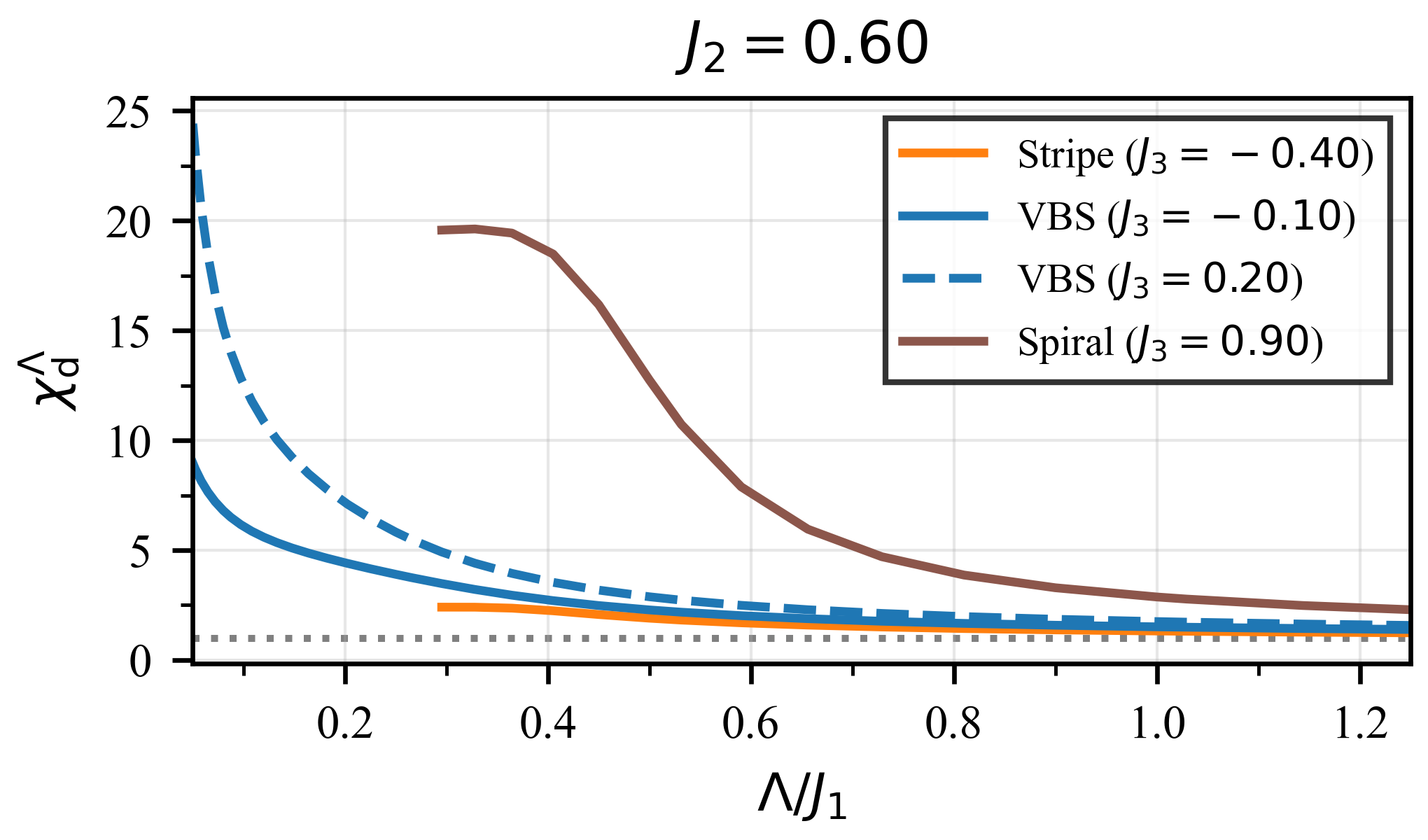}
    \caption{The flow of the dimer susceptibility $\chi_d^\Lambda$ for selected points.
    The gray dotted line indicates the unperturbed value $\chi_d=1$. The upper panel shows the $J_2=0$ cut. The candidate QSL regime ($J_3=0.35$, red) is characterized by a suppression of $\chi_d^\Lambda$ towards the end of the flow. In contrast, the VBS phase ($J_3=0.6$, blue) exhibits a steady, continuous enhancement. The AFM ($J_3=0.2$, green) and spiral ($J_3=1.0$, brown) phases undergo a magnetic flow breakdown at large scale $\Lambda$. The lower panel shows the $J_2=0.6$ cut. The system transitions from a stripe phase ($J_3=-0.4$, orange) into an extended VBS phase ($J_3=-0.1$ blue solid line and $J_3=0.2$ blue dotted line) characterized by continuous growth, and finally enters into a spiral phase ($J_3=0.9$, brown).}
    \label{fig:combined_VBS_flows}
\end{figure} 

Having mapped the phase boundaries of the quantum paramagnetic regime in the phase diagram, we apply the bond-anisotropy analysis outlined in Sec.~\ref{sec:j1j2} to track the dimer susceptibility $\chi^\Lambda_{p,d}$ of Eq.~\eqref{eq:order}. We take various cuts (see Fig.\ref{fig:phase_diagram}) throughout the entire QP region using $2\times1$ as well as $2\times2$ unit cells. In all cases, we observe that both unit-cell choices yield the same results for $\chi^\Lambda_{p}$ and $\chi^\Lambda_{d}$, indicating that pf-fRG cannot distinguish a columnar dimer pattern from a plaquette pattern, as previously reported by Reuther and Wölfle~\cite{reutherJ1J2Frustrated2010}.

To distinguish a VBS from a paramagnetic phase with no spatial symmetry breaking (e.g. a candidate quantum spin liquid), we systematically track the evolution of the dimer susceptibility $\chi_{p,d}^\Lambda$ throughout the RG flow. While $\chi^\Lambda_{p,d}$ generally increases during the initial stages of the flow, it develops rather distinctive behaviors as the sliding RG scale $\Lambda$ is reduced further. For certain points within the QP region, $\chi^\Lambda_{p,d}$ grows continuously down to the infrared; we take this as evidence for the formation of VBS order. In contrast, for other points, $\chi^\Lambda_{p,d}$ turns over at a finite scale and decreases near the end of the flow, $\Lambda \to 0$; such suppressed dimer response indicates the absence of VBS order.

To illustrate these contrasting behaviors, we analyze the flow of $\chi_{d}^\Lambda$ along two representative cuts, $J_2=0$ and $J_2=0.6$, shown in the upper and lower panels of Fig.~\ref{fig:combined_VBS_flows}, respectively. Along the $J_2=0$ cut, the red curve at the representative point $J_3=0.35$ shows the characteristic suppression of the susceptibility at low 
frequency scale to indicate a candidate QSL. This contrasts sharply with the blue curve ($J_3=0.6$) which exhibits continuous enhancement of $\chi_{d}^\Lambda$ to indicate VBS. For completeness, we also depict the neighboring AFM state (green curve, $J_3=0.2$); here, the dimer susceptibility is simply suppressed before the flow ultimately breaks down due to the magnetic instability. The second cut along $J_2=0.6$ does not intersect a candidate QSL region. Accordingly, the dimer susceptibility grows continuously within the VBS phase (depicted by the solid and dashed blue lines for $J_3=-0.1$ and $J_3=0.2$, respectively). This example further illustrates the key difference between QP and phases with long-range order, such as the stripe (orange curve, $J_3=-0.4$) or spiral (brown curve, $J_3=0.9$) order, whose flows breaks down at a finite critical scale $\Lambda_c$ to mark the magnetic instabilities.
 
By analyzing the flow of dimer susceptibility for each point within the QP region, we are able to further resolve it into two distinct regions: a candidate QSL regime (blue dots, shaded by red lines) and a likely VBS regime (blue dots only, unshaded) as shown in Fig.~\ref{fig:phase_diagram}. 
The robustness of this classification against the spatial truncation range is presented in \cite{supplementary_materials}.

Within the QP window, PEPS resolves a plaquette VBS and a mixed columnar-plaquette VBS~\cite{liuTensorNetworkStudy2024}. This distinction lies beyond the reach of our present analysis based on the dimer and plaquette susceptibilities. The designation of a single VBS region reported in our work should be understood as encompassing both possibilities. Comparing our global phase boundaries to recent tensor network studies~\cite{liuEmergenceGaplessQuantum2022,liuTensorNetworkStudy2024} reveals a further distinction: our pf-fRG calculations predict a significantly more extensive candidate QSL region, which extends deeply into the upper-left quadrant of the phase diagram. This discrepancy highlights the inherent biases of the two numerical techniques. In PEPS, the entanglement entropy is bounded by its finite bond dimension, which limits its capability to represent a gapless QSL. The pf-fRG approach is known to overestimate the destructive effect of quantum fluctuations on ordered states, thereby favoring an extensive disordered paramagnetic window. The two methods are therefore biased in opposite directions, so that the physical boundary likely lies between the two estimates.

\section{\label{sec:level6}Conclusion}

In this work, we have revisited the $J_1$--$J_2$--$J_3$ Heisenberg model on the square lattice using pf-fRG. 
The main result, the global phase diagram Fig.~\ref{fig:phase_diagram}, substantially extended the parameter regions, resolved and refined the phase boundaries achieved in the earlier pf-fRG study~\cite{reutherQuantumPhasesPlanar2011}.

In particular, our investigation covered negative values of both $J_2$ and $J_3$, and evaluated the dimer and plaquette response of this model for the first time. These technical advancements owe partly to the performance offered by the \texttt{PFFRGSolver.jl} package~\cite{kieseMultiloopFunctionalRenormalization2022}: its adaptive frequency grids and asymptotic classes improve the vertex resolution, and it allows larger lattice truncation ranges than the earlier study. The standard $C_4$-symmetric pf-fRG calculations capture the boundaries of the long-range ordered magnetic phases well, but overestimate the extent of the QP regime compared to  recent tensor network results~\cite{liuTensorNetworkStudy2024,liuEmergenceGaplessQuantum2022}.
We confirm that quantum fluctuations melt the classical $(\pi,q)$ spiral, leaving $\mathbf{q}=(q,q)$ as the only stabilized spiral mode, in agreement with the earlier pf-fRG study~\cite{reutherQuantumPhasesPlanar2011} and with PEPS~\cite{liuTensorNetworkStudy2024}. We further locate the point where the AFM, stripe, and QP regions meet near $J_2 \approx 0.55 J_1$ and $J_3$ between $-0.20 J_1$ and $-0.15J_1$. In PEPS this region closes far sooner, already for $J_3 \lesssim -0.05 J_1$~\cite{liuTensorNetworkStudy2024}. 

To probe the nature of the extended QP region, we track the dimer susceptibility defined in Eq.~\eqref{eq:order}. This probe was introduced for the $J_1$--$J_2$ model~\cite{reutherJ1J2Frustrated2010} but has not previously been applied to the $J_1$--$J_2$--$J_3$ model~\cite{reutherQuantumPhasesPlanar2011}. Consistent with the results of the $J_1$--$J_2$ pf-fRG study~\cite{reutherJ1J2Frustrated2010}, we find no qualitative difference in the susceptibility response between the plaquette and dimer susceptibilities. In Ref.~\cite{reutherJ1J2Frustrated2010}, the dimer and plaquette susceptibilities were enhanced at the end of the flow without developing a divergence, leaving spin liquid and VBS phase possibility equally open. 
Our calculation here, however, reveals a clear distinction in the low-energy behavior of these susceptibilities. This distinction separates the QP window into two regions: one in which $\chi_{p,d}^\Lambda$ grows continuously down to the lowest renormalization scale, which we take as supporting VBS order, and one in which it reaches a peak at a finite scale and is subsequently suppressed, leaving room for a candidate QSL. Because the two unit-cell choices give the same response, the precise nature of the VBS order remains elusive. Future work is needed to elucidate the VBS region reported here to compare with the two distinct valence bond solids predicted by PEPS~\cite{liuTensorNetworkStudy2024}. We emphasize that the absence of an enhanced dimer susceptibility does not by itself establish a spin liquid ground state. Settling this question within pf-fRG would require access to the higher-order correlation functions that lie beyond the present truncation scheme.

\begin{acknowledgments}
This work is supported by the Scientific and Technological Research Council of Turkiye (TUBITAK)
1001 program Project No. 124F122 (AK). The numerical calculations reported in this paper were partially performed using the High Performance and Grid Computing Center (TRUBA resources) at TUBITAK ULAKBIM, and the EuroHPC Joint Undertaking (EuroHPC JU) supercomputer MareNostrum 5, hosted by the Barcelona Supercomputing Center (BSC). Access to MareNostrum 5 was granted through a national access call coordinated by the Scientific and Technological Research Council of Turkey (TÜBİTAK). We gratefully acknowledge BSC, TÜBİTAK, and the EuroHPC JU for providing access to these resources and supporting this research. E.Z. acknowledges support from AFOSR Grant FA9550-23-1-0598 and NSF Grant PHY-206419.
\end{acknowledgments}

\section*{Data Availability}

The data that support the findings of this article are openly available in Ref.~\cite{gunalEvidenceValenceBond2026}.

\bibliographystyle{apsrev4-2}
\bibliography{references}

@misc{gunalEvidenceValenceBond2026,
	title = {Evidence of valence bond solid and quantum paramagnetism in the \${J}\_1\$--\${J}\_2\$--\${J}\_3\$ {Heisenberg} model from pseudofermion functional renormalization group},
	copyright = {Creative Commons Attribution 4.0 International},
	url = {https://zenodo.org/doi/10.5281/zenodo.23469635},
	doi = {10.5281/ZENODO.23469635},
	urldate = {2026-10-19},
	publisher = {Zenodo},
	author = {GÜNAL, GÜN and Zhao, Erhai and Keles, Ahmet},
	month = oct,
	year = {2026},
}

@misc{supplementary_materials,
	title = {Supplemental {Material}},
	note = {See Supplemental Material at [URL will be inserted by publisher] for the implementation of lattice symmetries and the dependence of the static and dimer susceptibility on the truncation range L.},
}

@article{balentsSpinLiquidsFrustrated2010,
	title = {Spin liquids in frustrated magnets},
	volume = {464},
	issn = {0028-0836, 1476-4687},
	url = {https://www.nature.com/articles/nature08917},
	doi = {10.1038/nature08917},
	language = {en},
	number = {7286},
	urldate = {2026-09-07},
	journal = {Nature},
	author = {Balents, Leon},
	month = mar,
	year = {2010},
	pages = {199--208},
}

@article{andersonResonatingValenceBonds1973,
	title = {Resonating valence bonds: {A} new kind of insulator?},
	shorttitle = {Resonating valence bonds},
	url = {https://sci-hub.ist/10.1016/0025-5408(73)90167-0},
	doi = {10.1016/0025-5408(73)90167-0},
	urldate = {2026-09-07},
	journal = {Materials Research Bulletin},
	author = {Anderson, P. W.},
	year = {1973},
}

@article{broholmQuantumSpinLiquids2020,
	title = {Quantum spin liquids},
	volume = {367},
	issn = {0036-8075, 1095-9203},
	url = {https://www.science.org/doi/10.1126/science.aay0668},
	doi = {10.1126/science.aay0668},
	language = {en},
	number = {6475},
	urldate = {2026-09-07},
	journal = {Science},
	author = {Broholm, C. and Cava, R. J. and Kivelson, S. A. and Nocera, D. G. and Norman, M. R. and Senthil, T.},
	month = jan,
	year = {2020},
	pages = {eaay0668},
}

@article{readValencebondSpinPeierlsGround1989,
	title = {Valence-bond and spin-{Peierls} ground states of low-dimensional quantum antiferromagnets},
	volume = {62},
	copyright = {http://link.aps.org/licenses/aps-default-license},
	issn = {0031-9007},
	url = {https://link.aps.org/doi/10.1103/PhysRevLett.62.1694},
	doi = {10.1103/PhysRevLett.62.1694},
	language = {en},
	number = {14},
	urldate = {2026-09-07},
	journal = {Physical Review Letters},
	author = {Read, N. and Sachdev, Subir},
	month = apr,
	year = {1989},
	pages = {1694--1697},
}

@article{andersonResonatingValenceBond1987,
	title = {The {Resonating} {Valence} {Bond} {State} in {La}\$\_\{2\}\${CuO}\$\_\{4\}\$ and {Superconductivity}},
	volume = {235},
	url = {http://www.jstor.org/stable/1698247},
	language = {en},
	number = {4793},
	journal = {Science, New Series},
	author = {Anderson, P. W.},
	year = {1987},
	pages = {1196--1198},
}

@article{ralkoGeneralizedQuantumDimer2009,
	title = {Generalized quantum dimer model applied to the frustrated {Heisenberg} model on the square lattice: {Emergence} of a mixed columnar-plaquette phase},
	volume = {80},
	copyright = {http://link.aps.org/licenses/aps-default-license},
	issn = {1098-0121, 1550-235X},
	shorttitle = {Generalized quantum dimer model applied to the frustrated {Heisenberg} model on the square lattice},
	url = {https://link.aps.org/doi/10.1103/PhysRevB.80.184427},
	doi = {10.1103/PhysRevB.80.184427},
	language = {en},
	number = {18},
	urldate = {2026-09-07},
	journal = {Physical Review B},
	author = {Ralko, A. and Mambrini, M. and Poilblanc, D.},
	month = nov,
	year = {2009},
	pages = {184427},
}

@article{arlegoPlaquetteOrder12008,
	title = {Plaquette order in the {J} 1 − {J} 2 − {J} 3 model: {Series} expansion analysis},
	volume = {78},
	copyright = {http://link.aps.org/licenses/aps-default-license},
	issn = {1098-0121, 1550-235X},
	shorttitle = {Plaquette order in the {J} 1 − {J} 2 − {J} 3 model},
	url = {https://link.aps.org/doi/10.1103/PhysRevB.78.224415},
	doi = {10.1103/PhysRevB.78.224415},
	language = {en},
	number = {22},
	urldate = {2026-09-07},
	journal = {Physical Review B},
	author = {Arlego, Marcelo and Brenig, Wolfram},
	month = dec,
	year = {2008},
	pages = {224415},
}

@article{leungNumericalEvidenceSpinPeierls1996,
	title = {Numerical evidence for the spin-{Peierls} state in the frustrated quantum antiferromagnet},
	volume = {53},
	copyright = {http://link.aps.org/licenses/aps-default-license},
	issn = {0163-1829, 1095-3795},
	url = {https://link.aps.org/doi/10.1103/PhysRevB.53.2213},
	doi = {10.1103/PhysRevB.53.2213},
	language = {en},
	number = {5},
	urldate = {2026-09-07},
	journal = {Physical Review B},
	author = {Leung, P. W. and Lam, Ngar-wing},
	month = feb,
	year = {1996},
	pages = {2213--2216},
}

@article{sindzingrePhaseDiagramSpin12009,
	title = {Phase diagram of the spin-1/2 \textit{{J}}$_{\textrm{1}}$ - \textit{{J}}$_{\textrm{2}}$ - \textit{{J}}$_{\textrm{3}}$ {Heisenberg} model on the square lattice with ferromagnetic \textit{{J}}$_{\textrm{1}}$},
	volume = {145},
	issn = {1742-6596},
	url = {https://iopscience.iop.org/article/10.1088/1742-6596/145/1/012048},
	doi = {10.1088/1742-6596/145/1/012048},
	urldate = {2026-09-07},
	journal = {Journal of Physics: Conference Series},
	author = {Sindzingre, Philippe and Seabra, L and Shannon, Nic and Momoi, Tsutomu},
	month = jan,
	year = {2009},
	pages = {012048},
}

@article{murgExploringFrustratedSpin2009,
	title = {Exploring frustrated spin systems using projected entangled pair states},
	volume = {79},
	copyright = {http://link.aps.org/licenses/aps-default-license},
	issn = {1098-0121, 1550-235X},
	url = {https://link.aps.org/doi/10.1103/PhysRevB.79.195119},
	doi = {10.1103/PhysRevB.79.195119},
	language = {en},
	number = {19},
	urldate = {2026-09-07},
	journal = {Physical Review B},
	author = {Murg, V. and Verstraete, F. and Cirac, J. I.},
	month = may,
	year = {2009},
	pages = {195119},
}

@article{ferrerSpinliquidPhaseFrustrated1993,
	title = {Spin-liquid phase for the frustrated quantum {Heisenberg} antiferromagnet on a square lattice},
	volume = {47},
	copyright = {http://link.aps.org/licenses/aps-default-license},
	issn = {0163-1829, 1095-3795},
	url = {https://link.aps.org/doi/10.1103/PhysRevB.47.8769},
	doi = {10.1103/PhysRevB.47.8769},
	language = {en},
	number = {14},
	urldate = {2026-08-08},
	journal = {Physical Review B},
	author = {Ferrer, Jaime},
	month = apr,
	year = {1993},
	pages = {8769--8782},
}

@article{chubukovFirstorderTransitionFrustrated1991,
	title = {First-order transition in frustrated quantum antiferromagnets},
	volume = {44},
	copyright = {http://link.aps.org/licenses/aps-default-license},
	issn = {0163-1829, 1095-3795},
	url = {https://link.aps.org/doi/10.1103/PhysRevB.44.392},
	doi = {10.1103/PhysRevB.44.392},
	language = {en},
	number = {1},
	urldate = {2026-08-08},
	journal = {Physical Review B},
	author = {Chubukov, Andrey},
	month = jul,
	year = {1991},
	pages = {392--394},
}

@article{moreoIncommensurateCorrelationsFrustrated1990,
	title = {Incommensurate correlations in the \textit{t} - \textit{{J}} and frustrated spin-1/2 {Heisenberg} models},
	volume = {42},
	copyright = {http://link.aps.org/licenses/aps-default-license},
	issn = {0163-1829, 1095-3795},
	url = {https://link.aps.org/doi/10.1103/PhysRevB.42.6283},
	doi = {10.1103/PhysRevB.42.6283},
	language = {en},
	number = {10},
	urldate = {2026-08-07},
	journal = {Physical Review B},
	author = {Moreo, Adriana and Dagotto, Elbio and Jolicoeur, Thierry and Riera, José},
	month = oct,
	year = {1990},
	pages = {6283--6293},
}

@misc{gresistaUnconventionalOrdersMapleleaf2025,
	title = {Unconventional orders in the maple-leaf ferro-antiferromagnetic {Heisenberg} model},
	url = {http://arxiv.org/abs/2511.21598},
	doi = {10.48550/arXiv.2511.21598},
	language = {en},
	urldate = {2026-02-23},
	publisher = {arXiv},
	author = {Gresista, Lasse and Kiese, Dominik and Trebst, Simon and Iqbal, Yasir},
	month = nov,
	year = {2025},
	note = {arXiv:2511.21598 [cond-mat]},
}

@article{capriottiLowTemperatureBrokenSymmetryPhases2004,
	title = {Low-{Temperature} {Broken}-{Symmetry} {Phases} of {Spiral} {Antiferromagnets}},
	volume = {93},
	copyright = {http://link.aps.org/licenses/aps-default-license},
	issn = {0031-9007, 1079-7114},
	url = {https://link.aps.org/doi/10.1103/PhysRevLett.93.257206},
	doi = {10.1103/PhysRevLett.93.257206},
	language = {en},
	number = {25},
	urldate = {2026-02-13},
	journal = {Physical Review Letters},
	author = {Capriotti, Luca and Sachdev, Subir},
	month = dec,
	year = {2004},
	pages = {257206},
}

@article{huangPlaquettetypeValenceBond2024,
	title = {Plaquette-type valence bond solid state in the {J} 1 − {J} 2 square-lattice {Heisenberg} model},
	volume = {110},
	issn = {2469-9950, 2469-9969},
	url = {https://link.aps.org/doi/10.1103/PhysRevB.110.195111},
	doi = {10.1103/PhysRevB.110.195111},
	language = {en},
	number = {19},
	urldate = {2025-08-08},
	journal = {Physical Review B},
	author = {Huang, Jiale and Qian, Xiangjian and Qin, Mingpu},
	month = nov,
	year = {2024},
	pages = {195111},
}

@article{wuPhaseDiagramMagnetic2022,
	title = {Phase diagram and magnetic excitations of {J} 1 – {J} 3 {Heisenberg} model on the square lattice},
	volume = {106},
	issn = {2469-9950, 2469-9969},
	url = {https://link.aps.org/doi/10.1103/PhysRevB.106.125129},
	doi = {10.1103/PhysRevB.106.125129},
	language = {en},
	number = {12},
	urldate = {2025-08-07},
	journal = {Physical Review B},
	author = {Wu, Muwei and Gong, Shou-Shu and Yao, Dao-Xin and Wu, Han-Qing},
	month = sep,
	year = {2022},
	pages = {125129},
}

@article{mullerPseudofermionFunctionalRenormalization2024,
	title = {Pseudo-fermion functional renormalization group for spin models},
	volume = {87},
	copyright = {http://creativecommons.org/licenses/by/4.0},
	issn = {0034-4885, 1361-6633},
	url = {https://iopscience.iop.org/article/10.1088/1361-6633/ad208c},
	doi = {10.1088/1361-6633/ad208c},
	language = {en},
	number = {3},
	urldate = {2025-08-03},
	journal = {Reports on Progress in Physics},
	publisher = {IOP Publishing},
	author = {Müller, Tobias and Kiese, Dominik and Niggemann, Nils and Sbierski, Björn and Reuther, Johannes and Trebst, Simon and Thomale, Ronny and Iqbal, Yasir},
	month = mar,
	year = {2024},
	pages = {036501},
}

@article{liuTensorNetworkStudy2024,
	title = {Tensor network study of the spin-12 square-lattice {J1}−{J2}−{J3} model: {Incommensurate} spiral order, mixed valence-bond solids, and multicritical points},
	volume = {109},
	copyright = {https://link.aps.org/licenses/aps-default-license},
	issn = {2469-9950, 2469-9969},
	shorttitle = {Tensor network study of the spin-12 square-lattice {J1}−{J2}−{J3} model},
	url = {https://link.aps.org/doi/10.1103/PhysRevB.109.235116},
	doi = {10.1103/physrevb.109.235116},
	language = {en},
	number = {23},
	urldate = {2025-08-01},
	journal = {Physical Review B},
	publisher = {American Physical Society (APS)},
	author = {Liu, Wen-Yuan and Poilblanc, Didier and Gong, Shou-Shu and Chen, Wei-Qiang and Gu, Zheng-Cheng},
	month = jun,
	year = {2024},
}

@article{iqbalIntertwinedNematicOrders2016,
	title = {Intertwined nematic orders in a frustrated ferromagnet},
	volume = {94},
	copyright = {http://link.aps.org/licenses/aps-default-license},
	issn = {2469-9950, 2469-9969},
	url = {https://link.aps.org/doi/10.1103/PhysRevB.94.224403},
	doi = {10.1103/physrevb.94.224403},
	language = {en},
	number = {22},
	urldate = {2025-07-31},
	journal = {Physical Review B},
	publisher = {American Physical Society (APS)},
	author = {Iqbal, Yasir and Ghosh, Pratyay and Narayanan, Rajesh and Kumar, Brijesh and Reuther, Johannes and Thomale, Ronny},
	month = dec,
	year = {2016},
}

@article{kiesePinchpointsHalfmoonsStars2023,
	title = {Pinch-points to half-moons and up in the stars: {The} kagome skymap},
	volume = {5},
	copyright = {https://creativecommons.org/licenses/by/4.0/},
	issn = {2643-1564},
	shorttitle = {Pinch-points to half-moons and up in the stars},
	url = {https://link.aps.org/doi/10.1103/PhysRevResearch.5.L012025},
	doi = {10.1103/physrevresearch.5.l012025},
	language = {en},
	number = {1},
	urldate = {2025-07-22},
	journal = {Physical Review Research},
	publisher = {American Physical Society (APS)},
	author = {Kiese, Dominik and Ferrari, Francesco and Astrakhantsev, Nikita and Niggemann, Nils and Ghosh, Pratyay and Müller, Tobias and Thomale, Ronny and Neupert, Titus and Reuther, Johannes and Gingras, Michel J. P. and Trebst, Simon and Iqbal, Yasir},
	month = feb,
	year = {2023},
}

@article{reutherQuantumPhasesPlanar2011,
	title = {Quantum phases of the planar antiferromagnetic {J} 1 − {J} 2 − {J} 3 {Heisenberg} model},
	volume = {83},
	copyright = {http://link.aps.org/licenses/aps-default-license},
	issn = {1098-0121, 1550-235X},
	url = {https://link.aps.org/doi/10.1103/PhysRevB.83.064416},
	doi = {10.1103/PhysRevB.83.064416},
	language = {en},
	number = {6},
	urldate = {2025-06-23},
	journal = {Physical Review B},
	author = {Reuther, Johannes and Wölfle, Peter and Darradi, Rachid and Brenig, Wolfram and Arlego, Marcelo and Richter, Johannes},
	month = feb,
	year = {2011},
	pages = {064416},
}

@article{liuGaplessQuantumSpin2022,
	title = {Gapless quantum spin liquid and global phase diagram of the spin-1/2 {J} 1 - {J} 2 square antiferromagnetic {Heisenberg} model},
	volume = {67},
	copyright = {https://www.elsevier.com/tdm/userlicense/1.0/},
	issn = {20959273},
	url = {https://linkinghub.elsevier.com/retrieve/pii/S2095927322001001},
	doi = {10.1016/j.scib.2022.03.010},
	language = {en},
	number = {10},
	urldate = {2025-06-20},
	journal = {Science Bulletin},
	author = {Liu, Wen-Yuan and Gong, Shou-Shu and Li, Yu-Bin and Poilblanc, Didier and Chen, Wei-Qiang and Gu, Zheng-Cheng},
	month = may,
	year = {2022},
	pages = {1034--1041},
}

@article{kieseMultiloopFunctionalRenormalization2022,
	title = {Multiloop functional renormalization group approach to quantum spin systems},
	volume = {4},
	issn = {2643-1564},
	url = {https://link.aps.org/doi/10.1103/PhysRevResearch.4.023185},
	doi = {10.1103/PhysRevResearch.4.023185},
	language = {en},
	number = {2},
	urldate = {2025-06-03},
	journal = {Physical Review Research},
	author = {Kiese, Dominik and Müller, Tobias and Iqbal, Yasir and Thomale, Ronny and Trebst, Simon},
	month = jun,
	year = {2022},
	pages = {023185},
}

@article{wangCriticalLevelCrossings2018,
	title = {Critical {Level} {Crossings} and {Gapless} {Spin} {Liquid} in the {Square}-{Lattice} {Spin}- 1 / 2 {J} 1 − {J} 2 {Heisenberg} {Antiferromagnet}},
	volume = {121},
	issn = {0031-9007, 1079-7114},
	url = {https://link.aps.org/doi/10.1103/PhysRevLett.121.107202},
	doi = {10.1103/PhysRevLett.121.107202},
	language = {en},
	number = {10},
	urldate = {2025-04-12},
	journal = {Physical Review Letters},
	author = {Wang, Ling and Sandvik, Anders W.},
	month = sep,
	year = {2018},
	pages = {107202},
}

@article{qianAbsenceSpinLiquid2024,
	title = {Absence of spin liquid phase in the {J} 1 − {J} 2 {Heisenberg} model on the square lattice},
	volume = {109},
	issn = {2469-9950, 2469-9969},
	url = {https://link.aps.org/doi/10.1103/PhysRevB.109.L161103},
	doi = {10.1103/PhysRevB.109.L161103},
	language = {en},
	number = {16},
	urldate = {2025-04-12},
	journal = {Physical Review B},
	author = {Qian, Xiangjian and Qin, Mingpu},
	month = apr,
	year = {2024},
	pages = {L161103},
}

@article{gottelCriticalScalesAnisotropic2012,
	title = {Critical scales in anisotropic spin systems from functional renormalization},
	volume = {85},
	copyright = {http://link.aps.org/licenses/aps-default-license},
	issn = {1098-0121, 1550-235X},
	url = {https://link.aps.org/doi/10.1103/PhysRevB.85.214406},
	doi = {10.1103/PhysRevB.85.214406},
	language = {en},
	number = {21},
	urldate = {2024-08-29},
	journal = {Physical Review B},
	author = {Göttel, Stefan and Andergassen, Sabine and Honerkamp, Carsten and Schuricht, Dirk and Wessel, Stefan},
	month = jun,
	year = {2012},
	pages = {214406},
}

@article{liuEmergenceGaplessQuantum2022,
	title = {Emergence of {Gapless} {Quantum} {Spin} {Liquid} from {Deconfined} {Quantum} {Critical} {Point}},
	volume = {12},
	issn = {2160-3308},
	url = {https://link.aps.org/doi/10.1103/PhysRevX.12.031039},
	doi = {10.1103/PhysRevX.12.031039},
	language = {en},
	number = {3},
	urldate = {2024-07-31},
	journal = {Physical Review X},
	author = {Liu, Wen-Yuan and Hasik, Juraj and Gong, Shou-Shu and Poilblanc, Didier and Chen, Wei-Qiang and Gu, Zheng-Cheng},
	month = sep,
	year = {2022},
	pages = {031039},
}

@article{iqbalFunctionalRenormalizationGroup2016,
	title = {Functional renormalization group for three-dimensional quantum magnetism},
	volume = {94},
	issn = {2469-9950, 2469-9969},
	url = {https://link.aps.org/doi/10.1103/PhysRevB.94.140408},
	doi = {10.1103/PhysRevB.94.140408},
	language = {en},
	number = {14},
	urldate = {2023-11-23},
	journal = {Physical Review B},
	author = {Iqbal, Yasir and Thomale, Ronny and Parisen Toldin, Francesco and Rachel, Stephan and Reuther, Johannes},
	month = oct,
	year = {2016},
	pages = {140408},
}

@article{reutherFunctionalRenormalizationGroup2011,
	title = {Functional renormalization group for the anisotropic triangular antiferromagnet},
	volume = {83},
	issn = {1098-0121, 1550-235X},
	url = {https://link.aps.org/doi/10.1103/PhysRevB.83.024402},
	doi = {10.1103/PhysRevB.83.024402},
	language = {en},
	number = {2},
	urldate = {2023-05-17},
	journal = {Physical Review B},
	author = {Reuther, Johannes and Thomale, Ronny},
	month = jan,
	year = {2011},
	pages = {024402},
}

@article{kataninFulfillmentWardIdentities2004,
	title = {Fulfillment of {Ward} identities in the functional renormalization group approach},
	volume = {70},
	issn = {1098-0121, 1550-235X},
	url = {https://link.aps.org/doi/10.1103/PhysRevB.70.115109},
	doi = {10.1103/PhysRevB.70.115109},
	language = {en},
	number = {11},
	urldate = {2023-05-04},
	journal = {Physical Review B},
	author = {Katanin, A. A.},
	month = sep,
	year = {2004},
	pages = {115109},
}

@article{jiangSpinLiquidGround2012,
	title = {Spin liquid ground state of the spin- 1/2 square {J1} - {J2} {Heisenberg} model},
	volume = {86},
	issn = {1098-0121, 1550-235X},
	url = {https://link.aps.org/doi/10.1103/PhysRevB.86.024424},
	doi = {10.1103/PhysRevB.86.024424},
	language = {en},
	number = {2},
	urldate = {2023-04-26},
	journal = {Physical Review B},
	author = {Jiang, Hong-Chen and Yao, Hong and Balents, Leon},
	month = jul,
	year = {2012},
	pages = {024424},
}

@article{richterSpin12J1J22010,
	title = {The spin-1/2 {J1}–{J2} {Heisenberg} antiferromagnet on the square lattice: {Exact} diagonalization for {N}=40 spins},
	volume = {73},
	issn = {1434-6028, 1434-6036},
	shorttitle = {The spin-1/2 {J1}–{J2} {Heisenberg} antiferromagnet on the square lattice},
	url = {http://link.springer.com/10.1140/epjb/e2009-00400-4},
	doi = {10.1140/epjb/e2009-00400-4},
	language = {en},
	number = {1},
	urldate = {2023-04-26},
	journal = {The European Physical Journal B},
	author = {Richter, J. and Schulenburg, J.},
	month = jan,
	year = {2010},
	pages = {117--124},
}

@article{gongPlaquetteOrderedPhase2014,
	title = {Plaquette {Ordered} {Phase} and {Quantum} {Phase} {Diagram} in the {Spin}- 1/2 {J1} − {J2} {Square} {Heisenberg} {Model}},
	volume = {113},
	issn = {0031-9007, 1079-7114},
	url = {https://link.aps.org/doi/10.1103/PhysRevLett.113.027201},
	doi = {10.1103/PhysRevLett.113.027201},
	language = {en},
	number = {2},
	urldate = {2023-04-26},
	journal = {Physical Review Letters},
	author = {Gong, Shou-Shu and Zhu, Wei and Sheng, D. N. and Motrunich, Olexei I. and Fisher, Matthew P. A.},
	month = jul,
	year = {2014},
	pages = {027201},
}

@article{wentzellHighfrequencyAsymptoticsVertex2020,
	title = {High-frequency asymptotics of the vertex function: {Diagrammatic} parametrization and algorithmic implementation},
	volume = {102},
	issn = {2469-9950, 2469-9969},
	shorttitle = {High-frequency asymptotics of the vertex function},
	url = {https://link.aps.org/doi/10.1103/PhysRevB.102.085106},
	doi = {10.1103/PhysRevB.102.085106},
	language = {en},
	number = {8},
	urldate = {2023-04-11},
	journal = {Physical Review B},
	author = {Wentzell, Nils and Li, Gang and Tagliavini, Agnese and Taranto, Ciro and Rohringer, Georg and Held, Karsten and Toschi, Alessandro and Andergassen, Sabine},
	month = aug,
	year = {2020},
	pages = {085106},
}

@article{kelesAbsenceLongrangeOrder2018,
	title = {Absence of long-range order in a triangular spin system with dipolar interactions},
	volume = {120},
	issn = {0031-9007, 1079-7114},
	doi = {10.1103/PhysRevLett.120.187202},
	language = {en},
	number = {18},
	urldate = {2022-09-10},
	journal = {Physical Review Letters},
	author = {Keleş, Ahmet and Zhao, Erhai},
	month = may,
	year = {2018},
	pages = {187202},
}

@article{iqbalSpinLiquidNature2016,
	title = {Spin liquid nature in the {Heisenberg} {J1} − {J2} triangular antiferromagnet},
	volume = {93},
	issn = {2469-9950, 2469-9969},
	url = {https://link.aps.org/doi/10.1103/PhysRevB.93.144411},
	doi = {10.1103/PhysRevB.93.144411},
	language = {en},
	number = {14},
	urldate = {2022-09-17},
	journal = {Physical Review B},
	author = {Iqbal, Yasir and Hu, Wen-Jun and Thomale, Ronny and Poilblanc, Didier and Becca, Federico},
	month = apr,
	year = {2016},
	pages = {144411},
}

@article{kelesRenormalizationGroupAnalysis2018,
	title = {Renormalization group analysis of dipolar {Heisenberg} model on square lattice},
	volume = {97},
	issn = {2469-9950, 2469-9969},
	url = {https://link.aps.org/doi/10.1103/PhysRevB.97.245105},
	doi = {10.1103/PhysRevB.97.245105},
	language = {en},
	number = {24},
	urldate = {2022-10-21},
	journal = {Physical Review B},
	author = {Keleş, Ahmet and Zhao, Erhai},
	month = jun,
	year = {2018},
	pages = {245105},
}

@misc{thoennissMultiloopPseudofermionFunctional2020,
	title = {Multiloop pseudofermion functional renormalization for quantum spin systems: {Application} to the spin-1/2 kagome {Heisenberg} model},
	shorttitle = {Multiloop pseudofermion functional renormalization for quantum spin systems},
	url = {http://arxiv.org/abs/2011.01268},
	language = {en},
	urldate = {2022-11-02},
	publisher = {arXiv:2011.01268},
	author = {Thoenniss, Julian and Ritter, Marc K. and Kugler, Fabian B. and von Delft, Jan and Punk, Matthias},
	month = nov,
	year = {2020},
	note = {arXiv:2011.01268 [cond-mat]},
}

@article{kelesRiseFallPlaquette2022,
	title = {Rise and fall of plaquette order in the {Shastry}-{Sutherland} magnet revealed by pseudofermion functional renormalization group},
	volume = {105},
	issn = {2469-9950, 2469-9969},
	url = {https://link.aps.org/doi/10.1103/PhysRevB.105.L041115},
	doi = {10.1103/PhysRevB.105.L041115},
	language = {en},
	number = {4},
	urldate = {2023-03-04},
	journal = {Physical Review B},
	author = {Keleş, Ahmet and Zhao, Erhai},
	month = jan,
	year = {2022},
	pages = {L041115},
}

@article{savaryQuantumSpinLiquids2017,
	title = {Quantum spin liquids: a review},
	volume = {80},
	issn = {0034-4885, 1361-6633},
	shorttitle = {Quantum spin liquids},
	url = {https://iopscience.iop.org/article/10.1088/0034-4885/80/1/016502},
	doi = {10.1088/0034-4885/80/1/016502},
	language = {en},
	number = {1},
	urldate = {2023-03-10},
	journal = {Reports on Progress in Physics},
	author = {Savary, Lucile and Balents, Leon},
	month = jan,
	year = {2017},
	pages = {016502},
}

@article{reutherJ1J2Frustrated2010,
	title = {J1− {J2} frustrated two-dimensional {Heisenberg} model: {Random} phase approximation and functional renormalization group},
	volume = {81},
	issn = {1098-0121, 1550-235X},
	shorttitle = {J1 − {J2} frustrated two-dimensional {Heisenberg} model},
	url = {https://link.aps.org/doi/10.1103/PhysRevB.81.144410},
	doi = {10.1103/PhysRevB.81.144410},
	language = {en},
	number = {14},
	urldate = {2023-01-23},
	journal = {Physical Review B},
	author = {Reuther, Johannes and Wölfle, Peter},
	month = apr,
	year = {2010},
	pages = {144410},
}

\clearpage
\pagebreak

\setcounter{equation}{0}
\setcounter{section}{0}
\setcounter{figure}{0}
\setcounter{table}{0}
\setcounter{page}{1}
\makeatletter
\renewcommand{\thefigure}{S\arabic{figure}}
\renewcommand{\thetable}{S\arabic{table}}
\renewcommand{\theequation}{S\arabic{equation}}
\renewcommand{\thesection}{\Roman{section}}   
\renewcommand{\thesubsection}{\thesection\alph{subsection}}   
\makeatother

\clearpage\phantomsection
\addcontentsline{toc}{chapter}{Supplementary Material}
\widetext
\begin{center}
  \maketitle
\textbf{\large Supplementary Materials for ''Evidence of valence bond solid and quantum paramagnetism in the $J_1$--$J_2$--$J_3$ Heisenberg model from pseudofermion functional renormalization group''}
\end{center}
\begin{center}
Gün Günal,
Erhai Zhao,
Ahmet Keleş
\end{center}

The SI includes the details of the implementation of lattice symmetries in two-particle vertex and examples to finite size scaling plots for magnetic susceptibility and dimer susceptibility.

\section{Implementation of Lattice Symmetries}\label{A:lattice}

The fundamental building blocks of the pf-fRG flow equations are the two-particle vertices, which are bilocal, i.e., they depend on only two lattice sites~\cite{reutherJ1J2Frustrated2010,mullerPseudofermionFunctionalRenormalization2024}. Translational invariance of the lattice allows us to place the first of these sites in the unit cell at the origin, and vertices belonging to pairs of sites that are related by a symmetry of the lattice are equal. We use both properties to significantly reduce the number of independent vertices kept in memory.

We assume an infinite lattice but keep only vertices whose two sites are separated by a Euclidean distance $|\mathbf{r}-\mathbf{r}'|\le L$ (in units of the lattice spacing); vertices beyond this truncation range are set to zero. Because our study involves three different unit-cell geometries (the square lattice with $C_4$ symmetry, a $2\times1$ dimer lattice, and a $2\times2$ plaquette lattice), we adopt a generalized sublattice notation. 

Each lattice site is specified by two indices: a Bravais lattice vector indicating its unit cell, and a basis index indicating the site within that cell. A general two-particle vertex is thus written as $\Gamma_{i,\alpha; j,\beta}$, where $i$ and $j$ denote the translation vectors of the respective unit cells, and the Greek letters $\alpha, \beta$ label the sublattice sites. 

To determine the minimal set of irreducible vertices, we apply a two-step symmetry reduction. First, using the translation operator $T_i$, every vertex is mapped such that its first site lies in the reference unit cell at the origin:
\begin{equation}
    T_i \Gamma_{i,\alpha; j,\beta} = \Gamma_{0,\alpha; j',\beta},
\end{equation}
where $j'=j-i$ is the relative unit-cell coordinate.

Second, if two vertices are related through a symmetry operation of the lattice, only one representative copy is kept in memory:
\begin{equation}
    P_k \Gamma_{0,\alpha; j,\beta} = \Gamma_{0,\tilde\alpha; \tilde\jmath,\tilde\beta},
\end{equation}
where $P_k$ denotes a symmetry operation of the lattice. 

\begin{figure}[b]
    \centering
    \includegraphics{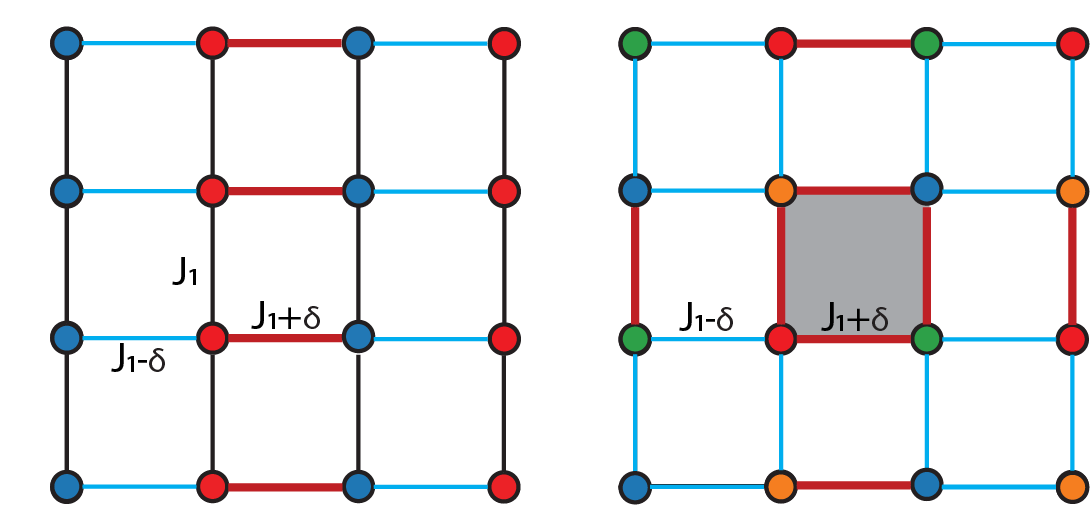}
    \caption{Enlarged unit cells used for the bond-anisotropy probe. Left: $2\times1$ dimer unit cell with red and blue sites. Right: $2\times2$ plaquette unit cell (shaded) with red, green, blue, and orange sites. Thick red bonds carry $J_1+\delta$, thin blue bonds $J_1-\delta$, and black bonds $J_1$.}
    \label{fig:unit_cells}
\end{figure}

For the square lattice, the symmetry reduction is straightforward. The unit cell contains only a single reference site ($\alpha = r$). By applying the symmetries of the square lattice, every pair of sites within the truncation range can be related to a symmetry-equivalent pair starting at the origin. For a truncation range of $L=7$, as an example, there are initially 149 sites within the truncation range of the origin; this procedure reduces them to just 25 irreducible sites.

For the dimer geometry, the bond anisotropy breaks the $C_4$ symmetry and enlarges the unit cell to $2\times1$. We label the two sites red and blue as illustrated in Fig.~\ref{fig:unit_cells} and choose the red site as our reference site. To find the irreducible vertices, we first determine the symmetry-reduced set for the red site. If a pair of sites starts at the blue site, after translating it to the origin unit cell we apply an additional reflection about the line parallel to the $y$ axis through the center of the unit cell. This reflection maps the blue site onto the red reference site, so that every vertex can be expressed in terms of vertices $\Gamma_{0,r;j,\beta}$. This procedure reduces the initial 149 sites to 82 irreducible sites for $L=7$.

For the plaquette geometry, the unit cell is enlarged to $2\times2$, containing four sites. We label these sites $\alpha \in \{r, g, b, o\}$ (red, green, blue, and orange), as illustrated in Fig.~\ref{fig:unit_cells}. We once again select the red site ($r$) as the reference site and compute all of its irreducible vertices. To evaluate vertices starting at the remaining sublattice sites, we exploit the $C_4$ rotational symmetry about the center of the plaquette: the $g$, $b$, and $o$ sites are mapped onto the $r$ site by a rotation of $90^\circ$ clockwise once, twice, or three times, respectively. The entire vertex structure is thus generated from the reference site, and the number of irreducible sites for $L=7$ is reduced to 79.

\section{Dependence on the Truncation Range}\label{A:fss}

\begin{figure}[h]
    \centering
     \includegraphics{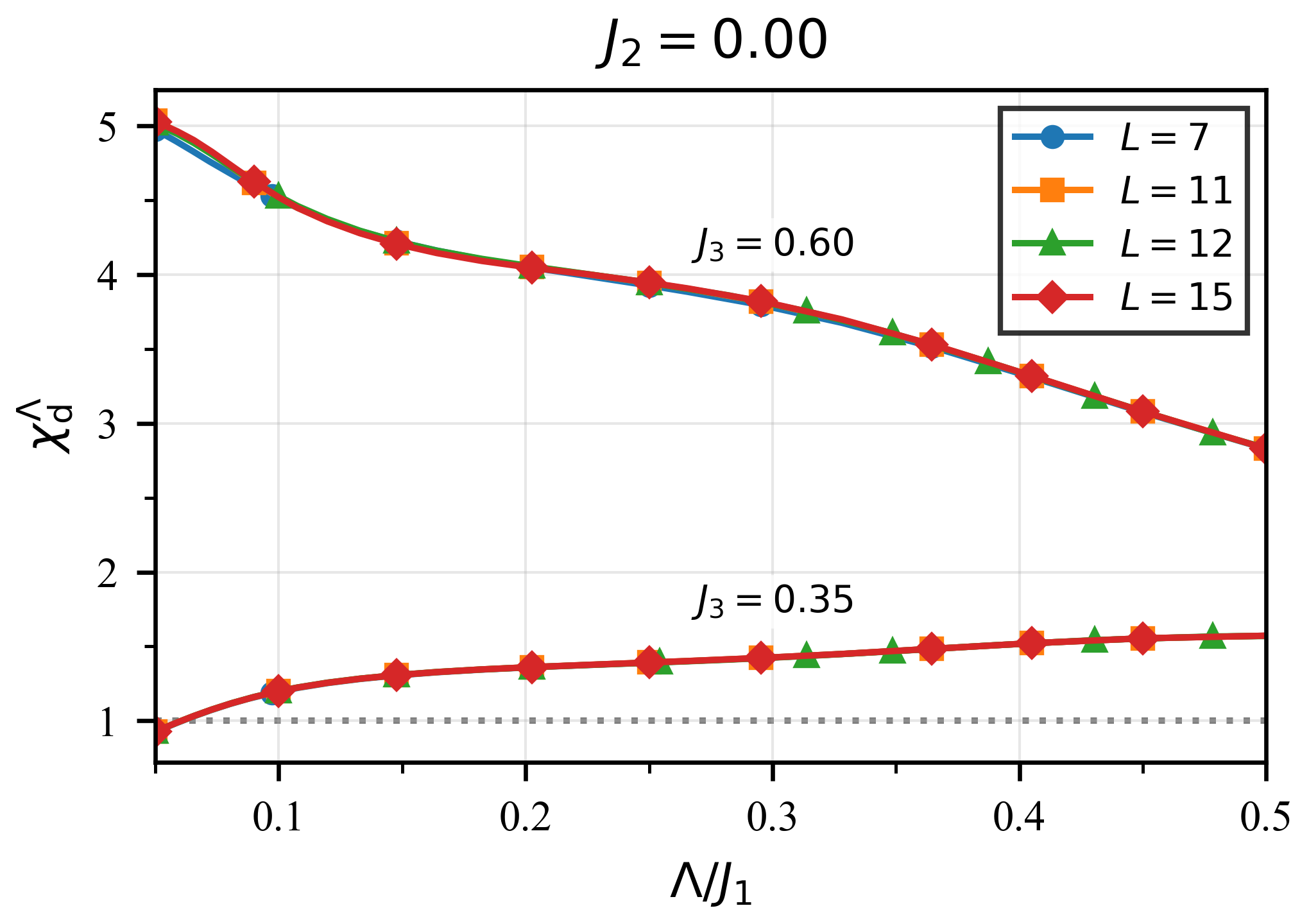}
    \caption{Dependence of the dimer susceptibility $\chi_d^\Lambda$ on the truncation range, for $L=7, 11, 12, 15$ along the $J_2=0$ cut for $J_3=0.35$ (candidate QSL regime) and $J_3=0.6$ (VBS phase). The characteristic low-energy scale suppression of $\chi_{d}$ is robust across all system sizes. The continuous enhancement of the susceptibility remains qualitatively unchanged with increasing lattice size in the VBS regime.}
    \label{fig:fss_sidebyside}
\end{figure}

To check the robustness of the classification introduced in Sec.~\ref{sec:j1j2}, we compute the flow of $\chi_d^\Lambda$ for truncation ranges $L=7, 11, 12, 15$. Figure~\ref{fig:fss_sidebyside} shows representative points in the candidate QSL ($J_3=0.35$) and VBS ($J_3=0.6$) regimes along the cut $J_2=0$. The overall flow structure remains qualitatively unchanged with increasing lattice size, in both the region where $\chi_d^\Lambda$ turns over and the region where it grows continuously.

\begin{figure}
    \centering
    \includegraphics[scale=0.6]{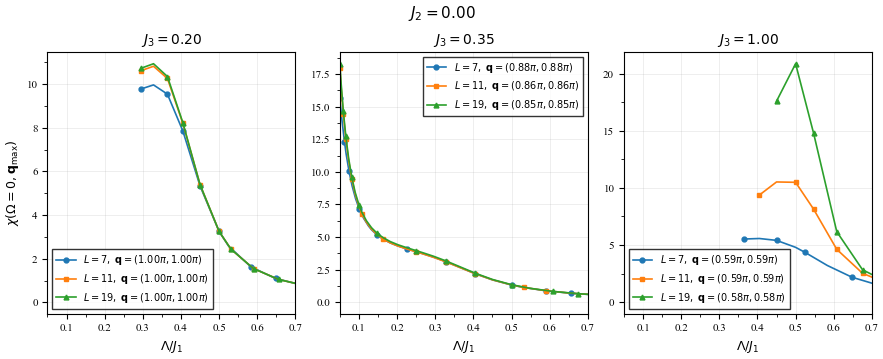}
    \caption{Flow of the static spin susceptibility at the leading wave vector, $\chi^\Lambda(\Omega=0,\mathbf{q}_{\max})$, along the $J_2=0$ cut for truncation ranges $L=7, 11, 19$. Left: AFM phase ($J_3=0.2$). Middle: QP regime ($J_3=0.35$). Right: spiral phase ($J_3=1.0$). In the ordered phases the flow breaks down at a finite scale $\Lambda_c$ and the peak height increases with $L$, most strongly in the spiral phase, whereas in the QP regime the flow stays smooth down to the lowest scale $\Lambda=0.05J_1$ for all $L$.}
    \label{fig:s3}
\end{figure}

\end{document}